\documentclass[longauth]{aa}

\newcommand{\feh}{\ensuremath{{\rm [Fe/H]}}}
\newcommand{\teff}{\ensuremath{T_{\rm eff}}}

\newcommand{\logg}{\ensuremath{\log{g}}}
\newcommand{\lightkurve}{\texttt{lightkurve}}
\newcommand{\astrocut}{\texttt{astrocut}}
\newcommand{\astroquery}{\texttt{astrocut}}
\newcommand{\tesseract}{\texttt{tesseract}}
\newcommand{\juliet}{\texttt{juliet}}
\newcommand{\zaspe}{\texttt{ZASPE}}
\newcommand{\ceres}{\texttt{CERES}}

\newcommand{\vsini}{\ensuremath{v \sin{i}}}

\newcommand{\mjup}{\ensuremath{{\rm M_{J}}}}
\newcommand{\mearth}{\ensuremath{{\rm M}_{\oplus}}}

\newcommand{\rjup}{\ensuremath{{\rm R_J}}}

  \newcommand{\mstA}   {\ensuremath{0.956 _{-0.051}^{+0.054}}}
  \newcommand{\rstA}   {\ensuremath{0.988 _{-0.012}^{+0.012}}}
  \newcommand{\rhostA} {\ensuremath{1.40   _{-0.14}^{+0.15}}}
  \newcommand{\lstA}   {\ensuremath{0.934 _{-0.031}^{+0.043}}}
  \newcommand{\ageA}   {\ensuremath{6.5   _{-2.1}^{+2.0}}}

  \newcommand{\mplA}   {\ensuremath{0.123\pm 0.012}}
  \newcommand{\rplA}   {\ensuremath{0.604\pm 0.028}}
  \newcommand{\teqA}   {\ensuremath{1086\pm 19}}

  \newcommand{\mstB}   {\ensuremath{0.972 _{-0.053}^{+0.054}}}
  \newcommand{\rstB}   {\ensuremath{1.086 _{-0.012}^{+0.013}}}
  \newcommand{\rhostB} {\ensuremath{1.07   _{-0.10}^{+0.11}}}
  \newcommand{\lstB}   {\ensuremath{1.113 _{-0.052}^{+0.054}}}
  \newcommand{\ageB}   {\ensuremath{8.4   _{-2.0}^{+2.0}}}

  \newcommand{\mplB}   {\ensuremath{0.213\pm 0.024}}
  \newcommand{\rplB}   {\ensuremath{0.991\pm 0.044}}
  \newcommand{\teqB}   {\ensuremath{1040\pm 19}}
  
  \newcommand{\teffA}   {\ensuremath{5697 \pm 80}}
  \newcommand{\loggA}   {\ensuremath{4.429 \pm 0.022}}
  \newcommand{\fehA}   {\ensuremath{0.02 \pm 0.04}}
  \newcommand{\vsiniA}   {\ensuremath{2.53 \pm 0.30}}
  
  \newcommand{\teffB}   {\ensuremath{5696 \pm 80}}
  \newcommand{\loggB}   {\ensuremath{4.355 \pm 0.022}}
  \newcommand{\fehB}   {\ensuremath{0.10 \pm 0.04}}
  \newcommand{\vsiniB}   {\ensuremath{3.65 \pm 0.30}}

\newcommand{\plnameA}{TOI-883b}
\newcommand{\stnameA}{TOI-883}
\newcommand{\plnameB}{TOI-899b}
\newcommand{\stnameB}{TOI-899}

\usepackage{graphicx}
\usepackage{txfonts}
\usepackage[colorlinks=true, citecolor=blue]{hyperref}
\usepackage[flushleft]{threeparttable}
\usepackage{booktabs}
\usepackage{caption}
\usepackage{subcaption}
\usepackage{longtable}
\usepackage{multirow}
\usepackage{siunitx}

\begin{document}

   \title{Two warm sub-Saturn mass planets identified from the TESS Full
Frame Images\thanks{This paper includes data gathered with the 6.5 meter Magellan Telescopes located at Las Campanas Observatory, Chile}}

   %\subtitle{I. Overviewing the $\kappa$-mechanism}

   \author{Felipe I. Rojas
          \inst{\ref{puc},\ref{mas}}%\fnmsep\thanks{E-mail: firojas@uc.cl}
          \and
          Rafael Brahm\inst{\ref{uai},\ref{mas},\ref{dataobs}}
          \and
          Andr\'es Jord\'an,$^{\ref{uai},\ref{mas},\ref{dataobs}}$ %andres.jordan@uai.cl
          \and
          N\'estor Espinoza$^{\ref{stsci}}$       % nespinoza@stsci.edu
          \and
          Thomas Henning$^{\ref{mpia}}$          % henning@mpia.de
          \and
          Jan Eberhardt$^{\ref{mpia}}$            % eberhardt@mpia.de
          \and
          Melissa J. Hobson$^{\ref{geneve}}$        % melihobson@gmail.com
          \and
          Martin Schlecker$^{\ref{arizona}}$        % schlecker@arizona.edu
          \and
          Marcelo Tala Pinto$^{\ref{uai}}$    % marcelo.tala@edu.uai.cl
          \and
          Trifon Trifonov$^{\ref{mpia},\ref{sofia},\ref{heidelberg}}$         % trifonov@mpia.de 
          \and
          Lyu Abe$^{\ref{oca}}$                         %lyu.abe@oca.eu
          \and
          Gaspar Bakos$^{\ref{princeton}}$            % gbakos@astro.princeton.edu
          \and
          Mauro Barbieri$^{\ref{uda}}$          % mauro.barbieri@uda.cl
          \and
          Khalid Barkaoui$^{\ref{iac},\ref{eap}, \ref{liege}}$    %khalid.barkaoui@uliege.be  
          \and
          Christopher J. Burke$^{\ref{kavli}}$                     %cjburke@mit.edu
          \and
          R. Paul Butler$^{\ref{epl}}$          % bluaper@gmail.com 
          \and
          Ilaria Carleo$^{\ref{iac},\ref{ull}}$           %ilariacarleo.astro@gmail.com
          \and
          Karen A. Collins$^{\ref{harvard}}$               %karen.collins@cfa.harvard.edu    
          \and
          Jeffrey D. Crane$^{\ref{carnegie}}$       % crane@carnegiescience.edu 
          \and
          Zoltan Csubry$^{\ref{princeton}}$           % zcsubry@astro.princeton.edu 
          \and
          Phil Evans$^{\ref{elsauce}}$ % phil@astrofizz.com
          \and
          Tristan Guillot$^{\ref{oca}}$                %tristan.guillot@oca.eu
          \and
          Chelsea X. Huang$^{\ref{queensland}}$              %Chelsea.Huang@usq.edu.au
          \and
          Jon M. Jenkins$^{\ref{nasa}}$                %jon.jenkins@nasa.gov
          \and
          Matias Jones$^{\ref{eso}}$           %mjones@eso.org
          \and
          Diana Kossakowski$^{\ref{mpia}}$       % kossakowski@mpia.de 
          \and
          David W. Latham$^{\ref{harvard}}$    %dlatham@cfa.harvard.edu
          \and
          Andrew W. Mann$^{\ref{unc}}$                 %awmann@unc.edu
          \and
          Djamel Mékarnia$^{\ref{oca}}$                 %mekarnia@oca.eu
          \and
          Maximiliano Moyano$^{\ref{ucn}}$     % mmoyano@ucn.cl 
          \and
          Sangeetha Nandakumar$^{\ref{uda}}$    % an.sangeetha@gmail.com 
          \and
          Hugh P. Osborn$^{\ref{bern}}$                   %hugh.osborn@unibe.ch
          \and
          George Ricker$^{\ref{kavli}}$        %grr@space.mit.edu
          \and
          David Rodriguez$^{\ref{stsci}}$              %drodriguez@stsci.edu
          \and
          Paula Sarkis$^{\ref{mpia}}$            % paula.sarkis.90@gmail.com
          \and
          Richard P. Schwarz$^{\ref{harvard}}$       %rpschwarz@comcast.net
          \and
          Sara Seager$^{\ref{kavli},\ref{eapmit},\ref{aamit} }$      %seager@mit.edu
          \and
          Ramotholo Sefako$^{\ref{saao}}$               %rrs@saao.ac.za
          \and
          Stephen Shectman$^{\ref{carnegie}}$       % shec@carnegiescience.edu
          \and
          Gregor Srdoc$^{\ref{kotizarovci}}$                  %gregorsrdoc@gmail.com 
          \and
          Stephanie Striegel$^{\ref{seti},\ref{nasa}}$              %stephanie.striegel@nasa.gov
          \and
          Vincent Suc$^{\ref{uai},\ref{elsauce}}$             % vincent.suc@uai.cl
          \and
          Johanna Teske$^{\ref{eap}}$          % jteske@carnegiescience.edu 
          \and
          Ian Thompson$^{\ref{eap}}$            % ian@carnegiescience.edu 
          \and
          Pascal Torres-Miranda$^{\ref{puc},\ref{mas}}$          % pjtorres1@uc.cl 
          \and
          Roland Vanderspek$^{\ref{kavli}}$    %roland@space.mit.edu
          \and
          Jos\'e Vines$^{\ref{ucn}}$           %  jose.vines.l@gmail.com
          \and
          Sharon X. Wang$^{\ref{tsinghua}}$             % sharonw@mail.tsinghua.edu.cn 
          \and
          Joshua N. Winn$^{\ref{princeton}}$           %jnwinn@princeton.edu,
          \and
          Carl Ziegler$^{\ref{sfasu}}$               %Carl.Ziegler@sfasu.edu 
}

   \institute{Instituto de Astrof\'isica, Facultad de F\'isica, Pontificia Universidad Cat\'olica de Chile, Av. Vicu\~{n}a Mackenna 4860, Santiago, Chile. \label{puc} \\
              \email{firojas@uc.cl}
        \and
            Millennium Institute of Astrophysics, Santiago, Chile. \label{mas} \\
        \and
            Facultad de Ingenier\'ia y Ciencias, Universidad Adolfo Ib\'a\~nez, Av.\ Diagonal las Torres 2640, Pe\~nalol\'en, Santiago, Chile \label{uai} \\
        \and
            Data Observatory Foundation, Chile \label{dataobs} \\
        \and 
            Space Telescope Science  Institute, 3700 San Martin Drive, Baltimore, MD 21218, USA \label{stsci} \\
        \and
            Max-Planck-Institut f\"ur Astronomie, K\"onigstuhl 17, 69117 Heidelberg, Germany \label{mpia} \\
        \and
            Observatoire de Genève, Département d'Astronomie, Université de Genève, Chemin Pegasi 51, 1290 Versoix, Switzerland \label{geneve} \\
        \and
            Steward Observatory and Department of Astronomy, The University of Arizona, Tucson, AZ 85721, USA \label{arizona} \\
        \and 
            Department of Astronomy, Sofia University ``St Kliment Ohridski'', 5 James Bourchier Blvd, BG-1164 Sofia, Bulgaria \label{sofia} \\
        \and 
            Landessternwarte, Zentrum f\"ur Astronomie der Universit\"at Heidelberg, K\"onigstuhl 12, D-69117 Heidelberg, Germany \label{heidelberg} \\
        \and
            Department   of   Astrophysical   Sciences,   Princeton  University, 4 Ivy Lane, Princeton, NJ 08544, USA \label{princeton} \\
        \and
            INCT, Universidad de Atacama, calle Copayapu 485, Copiap\'o, Atacama, Chile \label{uda} \\
        \and
            Department of Physics and Kavli Institute for Astrophysics and Space Research, Massachusetts Institute of Technology, Cambridge, MA 02139, USA \label{kavli} \\
        \and
            Earth and Planets Laboratory, Carnegie Institution for Science, 5241 Broad Branch Road, NW, Washington, DC 20015, USA \label{epl} \\
        \and
            Instituto de Astrof\'isica de Canarias (IAC), Calle V\'ia L\'actea s/n, 38200, La Laguna, Tenerife, Spain \label{iac} \\
        \and
            Departamento de Astrof\'isica, Universidad de La Laguna (ULL), 38206 La Laguna, Tenerife, Spain \label{ull} \\
        \and
            Center for Astrophysics \textbar \ Harvard \& Smithsonian, 60 Garden St, Cambridge, MA 02138, USA \label{harvard} \\
        \and 
            Department of Earth, Atmospheric and Planetary Sciences, Massachusetts Institute of Technology, Cambridge, MA 02139, USA \label{eap} \\
        \and
            Astrobiology Research Unit, Universit\'e de Li\`ege, 19C All\'ee du 6 Ao\^ut, 4000 Li\`ege, Belgium \label{liege} \\
        \and 
            Kotizarovci Observatory, Sarsoni 90, 51216 Viskovo, Croatia  \label{kotizarovci} \\
        \and 
            South African Astronomical Observatory, P.O. Box 9, Observatory, Cape Town 7935, South Africa \label{saao} \\
        \and 
            Universit\'e C\^ote d'Azur, Observatoire de la C\^ote d'Azur, CNRS, Laboratoire Lagrange, Bd de l'Observatoire, CS 34229, 06304 Nice cedex 4, France \label{oca} \\
        \and 
            Observatories of the Carnegie Institution for Science, 813 Santa Barbara Street, Pasadena, CA 91101 \label{carnegie}  \\
        \and 
            El Sauce Observatory, Coquimbo Province, Chile \label{elsauce} \\
        \and 
            University of Southern Queensland, Centre for Astrophysics, West Street, Toowoomba, QLD 4350 Australia \label{queensland} \\
        \and 
            European Southern Observatory (ESO), Alonso de C\'ordova 3107, Vitacura, Casilla 19001, Santiago de Chile \label{eso} \\
        \and 
            NASA Ames Research Center, Moffett Field, CA 94035, USA \label{nasa}  \\
        \and 
            Department of Physics and Astronomy, The University of North Carolina at Chapel Hill, Chapel Hill, NC 27599, USA \label{unc} \\
        \and 
            Instituto de Astronom\'ia, Universidad Cat\'olica del Norte, Angamos 0610, 1270709 Antofagasta, Chile \label{ucn}  \\
        \and 
            Physikalisches Institut, University of Bern, Gesellsschaftstrasse 6, 3012 Bern, Switzerland \label{bern} \\
        \and 
            Department of Earth, Atmospheric and Planetary Sciences, Massachusetts Institute of Technology, Cambridge, MA 02139, USA \label{eapmit} \\
        \and 
            Department of Aeronautics and Astronautics, MIT, 77 Massachusetts Avenue, Cambridge, MA 02139, USA \label{aamit} \\
        \and 
            SETI Institute, Mountain View, CA 94043 USA \label{seti} \\
        \and 
            Department of Astronomy, Tsinghua University, Beijing 100084, China \label{tsinghua} \\
        \and 
            Department of Physics, Engineering and Astronomy, Stephen F. Austin State University, 1936 North Street, Nacogdoches, TX 75962 \label{sfasu} \\
             }

   \date{Received XXXX; accepted YYYY}

% \abstract{}{}{}{}{} 
% 5 {} token are mandatory
 
  \abstract
  % context heading (optional)
  % {} leave it empty if necessary  
   {Characterization of warm giants is crucial to constrain giant planet formation and evolution. Measuring the mass and radius of these planets, combined with their moderated irradiation, allows us to estimate their planetary bulk composition, which is a key quantity to comprehend giant planet formation and structure.}
  % aims heading (mandatory)
   {We present the discovery of two transiting warm giant planets orbiting solar-type stars from the
Transiting Exoplanet Survey Satellite (TESS), which were characterized by further spectroscopic and photometric ground-based observations. }
  % methods heading (mandatory)
   {We performed a joint analysis of photometric data with radial velocities to confirm and characterize \plnameA\ and \plnameB , two sub-Saturns orbiting solar-like stars. }
  % results heading (mandatory)
   {\plnameA\ and \plnameB\ have masses of \mplA\ \mjup\ and \mplB\ \mjup, radius of \rplA\ \rjup\ and \rplB\ \rjup , periods of 10.06 d and 12.85 d and equilibrium temperature of \teqA K and \teqB , respectively.}
  % conclusions heading (optional), leave it empty if necessary 
   {While having similar masses, orbital periods and stellar host properties, these planets seem to have different internal compositions, which could point to distinct formation histories. Both planets are suitable targets for atmospheric studies to further constrain formation scenarios of planets in the Neptune-Saturn mass range.}


   \maketitle
%
%-------------------------------------------------------------------

\section{Introduction}
\label{sec:intro}

The study of transiting planets having masses in the Neptune-Saturn range (sub-Saturns) is of particular interest in the context of giant planet formation because they can give us insights into some key processes involved in the core accretion scenario \citep{perri:1974,pollack:96,mizuno:1980}. Following this classical model, it is not easy to explain how these relatively heavy objects avoid triggering the runaway accretion of the surrounding gas to form massive gas giants \citep{rafikov:2006}. More recent variations of the core accretion model that invoke the formation of giant planets through the accretion of icy pebbles \citep{lambrechts:2012} allow for the formation of giant planets with large core-to-envelope mass ratios in the outer regions of the disk ($\sim$25 AU) but also closer to the water ice line \citep[e.g.][]{bitsch:2015}. In this scenario, the more massive envelope-dominated gas giant planets are formed between these two regions. The measurement of the metallicity and composition of transiting sub-Saturns is key for identifying if these planets could have been formed in the regions predicted by these or other theoretical models \citep[e.g.][]{mordasini:2012}.

Heavy element enrichment for transiting planets can be measured by comparing the measured planet masses and radii with predictions from structural models, which gives information about the bulk metallicity of the planet \citep[e.g.][]{mordasini:2012,thorngren:2016}. It has also been proposed that the determination of the atmospheric enrichment of transiting giant planets can be used to constrain their formation conditions \citep[e.g.][]{mordasini:2016,espinoza:2017,madhusudhan:2017}. In recent years, transiting sub-Saturns have attracted attention due to the detection of water features in their transmission spectra \citep[e.g.][]{wakeford:2018,kreidberg:2018,colon:2020}, which is compatible with the formation of these planets in the ice-rich regions of the protoplanetary disk.

Despite the recent encouraging results for this population of transiting planets, one of the main limitations for using the current sample of well characterized (planet masses and radii measured with a precision of 20\% or better) transiting sub-Saturns to infer properties about their structures and compositions
through the modeling of their interiors, is that most of them orbit extremely close to their parent stars. The structure of the more massive sub-Saturns may be subject to the same unknown mechanism that inflates the radius of hot Jupiters \citep{laughlin:2011}, while on the low mass end, highly irradiated hot Neptunes are expected to lose a significant fraction of their envelopes through photo-evaporation \citep{owen:2018}. Most of these short period sub-Saturns have been discovered with ground-based photometric surveys \citep[e.g.][]{bakos:2010,bakos:2015,pepper:2017,hellier:2019} that lack the required duty cycle to efficiently detect giant planets with periods longer than 10 days. The NASA \textit{Kepler} mission \citep{Kepler} and its \textit{K2} successor were responsible for finding the first dozen of transiting warm sub-Saturns whose masses were determined either from transiting timing variations (TTVs) of multiple systems \citep[e.g.][]{holman:2010,panichi:2018} and/or via radial velocity follow-up for those having brighter host stars \citep[e.g.][]{barragan:2016,brady:2018,vaneylen:2018,brahm:2019}. More recently, the TESS mission \citep{TESS} has started to efficiently detect transiting giant planet candidates with periods longer than 10 days, some of which have been found to have masses in the sub-Saturn regime \citep[e.g.][]{huber:2019,addison:2020,dalba:2020,diaz:2020}.

In this study, we present the discovery of two new transiting warm giant planets with sub-Saturn masses that were first identified as candidates from the TESS Full Frame Images (FFIs). These planets were discovered in the context of the Warm gIaNts with tEss (WINE) collaboration, which focuses on the systematic confirmation and orbital characterization of TESS transiting giant planets with orbital periods larger than 10 days \citep[e.g.][]{brahm:20192,jordan:2020,brahm:2020,schlecker:2020,hobson:2021,trifonov:2021,trifonov:2023,bozhilov:2023,brahm:2023,hobson:2023,eberhardt:2023,jones:2024}%\citep[e.g.][]{hd1397, wine1, wine2, wine3}.

\section{TESS data}
\subsection{TESS observations}
\stnameA\ was monitored by TESS from December 15, 2018, to January 6, 2019, in Sector 6 with a cadence of 30 minutes. It was subsequently observed in Sector 33 with a cadence of 2 minutes from December 18, 2020, to January 13, 2021.

On the other hand, \stnameB\ was monitored by TESS in its 13 southern sectors, spanning a year of semi-continuous observations from July 25, 2018 to July 17, 2019.
Besides having FFI data (30-minute cadence) for all these sectors, \stnameB\ has 2-minute cadence observations for sectors 9-13, 29, 31, 33-37, 39, 61, 62, 64-66 and 68-69.

The image data were reduced and analyzed using the Science Processing Operations Center (SPOC) pipeline \citep{Jenkins2016} at NASA Ames Research Center.

\label{sec:tess}
\subsection{Tesseract pipeline}
Both systems were detected on light curves extracted from the TESS FFIs. This was done using a pipeline developed by our team, called \tesseract\footnote{https://github.com/astrofelipe/tesseract}. 

\tesseract\ has two ways to access data from the FFIs. The first one and used in this work depends on TESScut\footnote{https://mast.stsci.edu/tesscut/}, a service that runs \astrocut\  \citep{astrocut} on the Mikulski Archive for Space Telescopes (MAST) servers, to download a cutout from TESS FFIs around the region of interest. The TESScut API is accessed through \astroquery\ \citep{astroquery}. The second mode involves accessing a local file that contains all the previously downloaded FFIs stacked together. The latter method was specially designed to generate large numbers of light curves, for example all bright targets ($T_{\mathrm{mag}} < 13$) from the TIC catalog. This avoids making an unreasonable number of calls to the TESScut service, reducing the time needed to generate a light curve. 

The next subsections describe the steps taken after accessing the data through any previous methods.

\subsubsection{Postage Stamps}
To obtain the light curve of a given target, the user can directly input the target's equatorial coordinates or the TIC ID. For the latter, its coordinates are resolved using \astroquery\ \citep{astroquery}. If local files are used, then a call to \texttt{tess-point}\footnote{https://github.com/christopherburke/tess-point} \citep{tesspoint} is made to obtain the Camera and CCD where the target falls to select the right FFI stack.

Knowing the coordinates, the World Coordinate System (WCS) data is used to find the pixel coordinates of the target along the FFI and a postage stamp of $21\times21$ pixels by default is created. This size is small enough to use a constant background estimation and at the same time, it gives a good number of pixels (441) to work with, considering that the aperture used for a 9th TESS Magnitude star tends to be between 6x6 (36) and 5x5 (25) pixels. Fainter targets are expected to cover an equal or smaller number of pixels. For this task, the \texttt{SExtractorBackground} function from \texttt{photutils} \citep{photutils} is used. Very bright ($T_{\mathrm{mag}} < 9$), faint ($T_{\mathrm{mag}} > 14$) targets, crowded fields and nonstellar targets are examples where the user must exercise caution with the selected aperture and background behavior.

\subsubsection{Aperture photometry}
By default, aperture selection on the cutouts follows the procedure of \cite{K2P2}. First, pixels above different flux thresholds are selected. These thresholds correspond to 1,3 and 5 times the median absolute deviation of the postage stamp flux. Over these groups, a Gaussian filter is applied and star masks are made employing the clustering algorithm DBSCAN \citep{DBSCAN} implemented in \texttt{scikit-learn} \citep{scikit-learn}. Finally, the watershed method, implemented in \texttt{scikit-image} \citep{scikit-image}, is applied to separate possible close targets on the same mask. The watershed method interprets the image as a topographic map and identifies boundaries associated with the watershed lines, as if water were flooding from the local minima. In this case, we simply use negative values of the flux in order to apply the method and select the sub-region that contains the already calculated pixel coordinate of the target.

Light curves are made for each aperture and stored in \lightkurve\ \citep{lightkurve} objects. These objects contain methods that allow us to flatten the light curves and calculate the Combined Differential Photometric Precision (CDPP) noise metric. The light curve with lower CDPP is selected.

The same procedure can be alternatively done using circular or custom apertures. These can be very useful in crowded fields and for targets near the saturation limit.

\subsection{Transit search}
An iterative transit search is made using Transit Least Squares \citep[TLS;][]{TLS} on all light curves for each sector. In particular, the TLS periodogram shows a significant signal of $P\sim 10$ d in \stnameA\ (Sector 6) and $P\sim 12$ d in \stnameB, which can be found in all Sectors from 1 to 13. A visual inspection of the phased light curves reveals a transit signal in both targets (Figures \ref{fig:phasedA} and \ref{fig:phasedB}). Then, all the in-transit points are masked and a new search is performed to search for additional transit signals. No additional signals were found. This procedure was also done using Box Least Squares \citep[BLS;][]{BLS} obtaining similar results.

When these targets were classified as planet candidate hosts, neither of them were classified as TOIs and all the data had to be extracted from FFIs, except for \stnameB\ in Sectors 9-13, proving the effectiveness of the pipeline.

\subsection{TOI status}
The SPOC conducted a transit search with an adaptive, noise-compensating matched filter \citep{Jenkins2002,Jenkins2010,Jenkins2020}, producing a threshold crossing event (TCE) for which an initial limb-darkened transit model was fitted \citep{Li2019} and a suite of diagnostic tests were conducted to help make or break the planetary nature of the signal \citep{Twicken2018}. The TESS Science Office (TSO) reviewed the vetting information and issued an alert for TOI-883.01 and TOI-899 b on July 12, 2019 and July 15, 2019 \citep{Guerrero2021} respectively. According to the difference image centroiding tests, the host star is located within $0.84 \pm 2.5$ arcsec of the transit signal source of TOI-883 b and within $0.16 \pm 2.7$ arcsec of the transit signal source of TOI-899 b. The Quick Look Pipeline \citep[QLP;][]{qlp1, qlp2} also identified and vetted the candidates in July, 2019.

\section{Ground-based follow-up}
\label{sec:ground}

\subsection{Speckle imaging and Gaia sources}
\begin{figure*}
    \centering
    \hfill\includegraphics[width=0.45\linewidth]{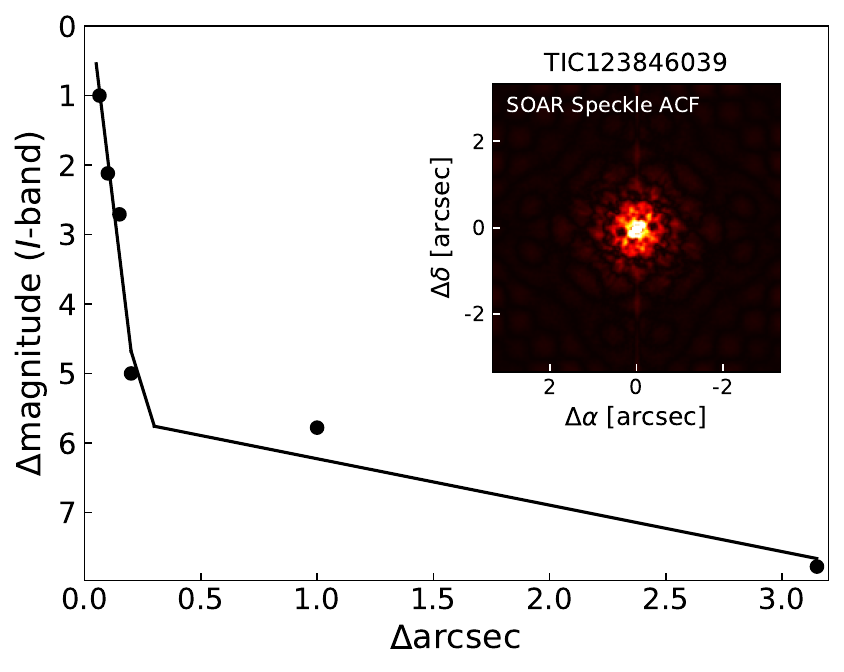}\hfill
    \includegraphics[width=0.45\linewidth]{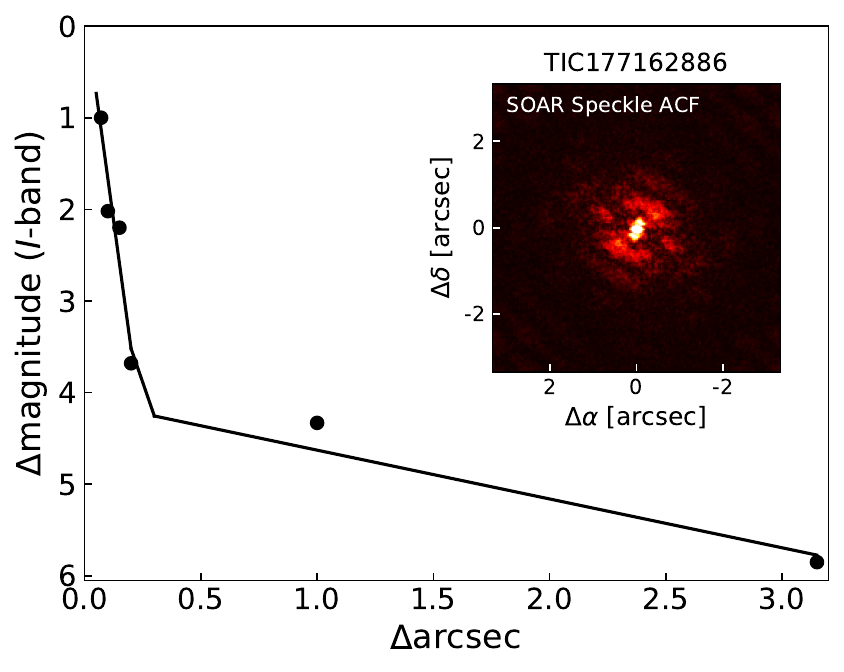}\hfill
    \caption{Speckle images for \stnameA\ and \stnameB\ obtained with the Southern Astrophysical Research (SOAR). Both observations show no evidence for close companions.}
    \label{fig:SOAR}
\end{figure*}

\stnameA\ and \stnameB\ were observed with the High-Resolution Camera (HRCam) installed at the 4.1m Southern Astrophysical Research \citep[SOAR,][]{SOAR} telescope located on Cerro Pach\'on, Chile, on November 9 of 2019, as part of the SOAR TESS Survey \citep{SOARTESS}. No nearby bright sources were identified for either target (Figure \ref{fig:SOAR}).

\begin{figure*}
    \centering
    \hfill\includegraphics[width=0.45\linewidth]{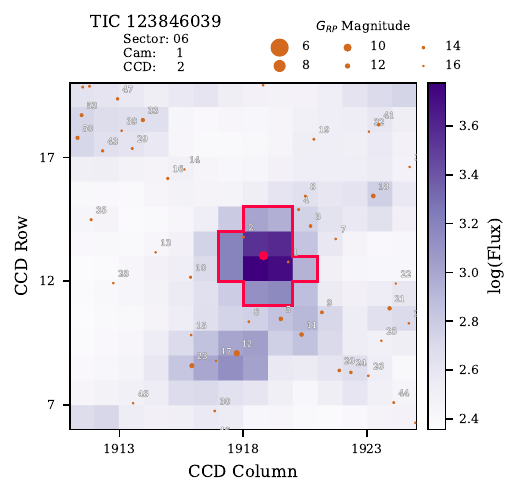}\hfill
    \includegraphics[width=0.45\linewidth]{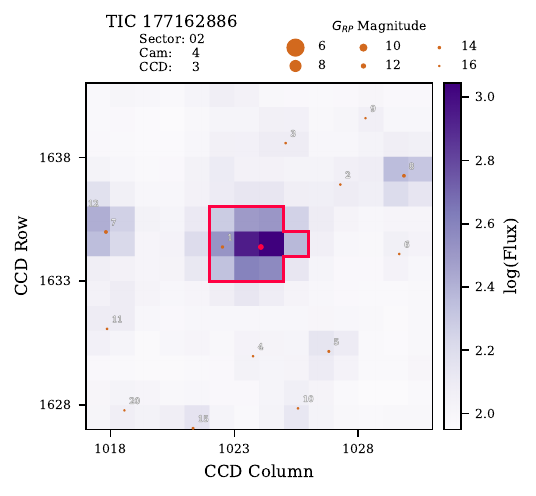}\hfill
    \caption{Postage stamps for \stnameA\ and \stnameB\ (Sector 2), respectively, taken from the TESS FFIs. Orange dots correspond to nearby Gaia sources, with marker size proportional to their magnitude. Target location and aperture used to generate the light curve are plotted in red.}
    \label{fig:Gaia}
\end{figure*}

Additionally, with \tesseract\ we generated plots including nearby Gaia \citep{gaiadr2summary} sources (up to $G_{RP}=17$), as seen in Figure \ref{fig:Gaia}). Both targets have companions, with \stnameA\ in a more crowded field but surrounded mostly by faint targets, showing a minimum of $\Delta T_{\mathrm{mag}} \approx 5.65$ inside the aperture, at $20''$, and \stnameB\ with a minimum of $\Delta T_{\mathrm{mag}} \approx 2.65$ inside the aperture, at $32''$.

Due to the TESS pixel scale ($21''$/pix), flux from nearby stars commonly falls inside the aperture used. To address this, we performed ground-based seeing-limited photometric follow-up observations with CHAT, El Sauce, LCOGT, and ASTEP to obtain a non-diluted transit depth, assuming the aperture is not contaminated by nearby sources. This is useful for estimating a dilution factor for TESS long cadence light curves, which will be used in the joint analysis. TESS short cadence light curves have been corrected for systematics, among other effects, including crowding, using the pipeline designed by the Science Processing Operations Center (SPOC). This correction is described in section 2.3.11 of \cite{pdcsap} and applied in the "PDCSAP" flux of these light curves.

\subsection{CHAT Photometry}
\stnameA\ was monitored over four separate nights using the 0.7-meter Chilean-Hungarian Automated Telescope (CHAT) located at Las Campanas Observatory. The observations were conducted in the Sloan \textit{i} passband, utilizing an exposure duration of 16 seconds and a slight defocus.

Due to the limited constraints on the candidate ephemeris from the TESS observations, only one of the CHAT light curves was able to confirm the presence of the transit. These observations occurred on the night of December 20 of 2019 and detected an egress of \plnameA\ during the first part of the observations.

\stnameB\ was also observed with CHAT on March 27 of 2019. A full transit was obtained in the Sloan \textit{i} passband. The adopted exposure time was 104 seconds and a mild defocus was applied which produced a typical peak flux of 15000 ADUs. Observations were interrupted for a couple of minutes after the ingress due to strong wind.

CHAT data for both targets were processed with a dedicated automated pipeline that was initially developed for data of the 1m LCOGT telescopes \citep{hats8}, and modified to produce light curves of CHAT in real time \citep[e.g.][]{mascara4,hats70,jones:2019,k2-287,hartman:2019}.

The transit obtained for \stnameA\ is displayed in Figure \ref{fig:phasedA} and for \stnameB\ in Figure \ref{fig:phasedB}.

\subsection{El Sauce Photometry}
\stnameB\ was observed from the Evans 0.36\,m telescope at El Sauce Observatory in Coquimbo Province, Chile. The telescope is equipped with a $1536\times1024$ SBIG STT-1603-3 camera. The image scale is 1$\farcs$47 pixel$^{-1}$, resulting in an $18.8\arcmin\times12.5\arcmin$ field of view. Observations were made using a Rc filter on February 20, 2020, covering a full transit, which is shown in Figure \ref{fig:phasedB}. The photometric data were extracted using {\tt AstroImageJ} \citep{Collins:2017} with circular $8\farcs$8 photometric apertures, which exclude all flux from all known neighboring stars in the Gaia DR3 catalog. 

\subsection{LCOGT}
\begin{figure}
    \centering
    \includegraphics[width=\hsize]{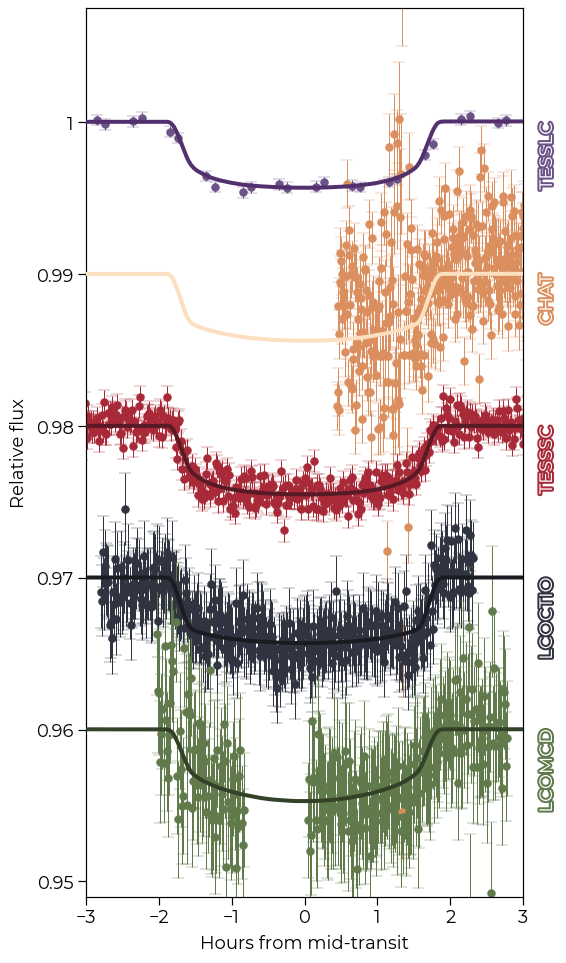}
    \caption{Photometric observations and transits for \plnameA. In the case of TESS, both light curves are phase folded. CHAT and both from Las Cumbres Observatory are follow-up data that observed a single transit.}
    \label{fig:phasedA}
\end{figure}

We used the Las Cumbres Observatory Global Telescope \citep[LCOGT;][]{Brown:2013} 1.0\,m network to observe both of our planet candidates. The 1.0\,m network $4096\times4096$ LCOGT SINISTRO cameras have an image scale of $0\farcs389$ per pixel, resulting in a $26\arcmin\times26\arcmin$ field of view. The images were calibrated by the standard LCOGT {\tt BANZAI} pipeline \citep{McCully:2018}, and photometric data were extracted using {\tt AstroImageJ}. \stnameA\ was observed on December 14, 2021, in the Sloan zs band simultaneously from both the Cerro Tololo Inter-American Observatory (CTIO) LCOGT node and the McDonald Observatory (McD) node. A full transit and a gapped full transit were obtained, respectively. The photometric data were extracted using circular $7\farcs$8 photometric apertures, which exclude all flux from all known neighboring stars in the Gaia DR3 catalog. The resulting lightcurves are shown in Figure \ref{fig:phasedA}. Full transits of \stnameB\ were observed on February 5, 2021, simultaneously in B and Sloan zs bands from the South Africa Astronomical Observatory (SAAO) node near Sutherland, South Africa. The photometric data were extracted using circular $5\farcs$4 photometric apertures, which exclude all flux from all known neighboring stars in the Gaia DR3 catalog. The resulting lightcurves are shown in Figure \ref{fig:phasedB}.

\subsection{ASTEP}
\stnameB\ was observed with the Antarctica Search for Transiting ExoPlanets (ASTEP) telescope installed at Concordia station on the East Antarctic plateau \citep{guillot2015,mekarnia2016}.

The 0.4 m telescope was equipped with an FLI Proline science camera with a KAF-16801E, $4096\times 4096$ front-illuminated CCD with an image scale of $0.93''$ pixel$^{-1}$ resulting in a $1\deg\times\deg$ corrected field of view. A dichroic plate was used to split the beam into a blue wavelength channel for guiding and a non-filtered red science channel roughly matching an Rc transmission curve \citep{Abe2013}. In January 2022, the focal box was replaced with a new one including two high sensitivity cameras for a two-colour (blue and red) simultaneous visible observations \citep{schmider2022}.

The telescope is automated or remotely operated when needed. Due to the extremely low data transmission rate at the Concordia Station, the data are automatically processed on-site. The raw light curves
are transferred to Europe on a server in Roma, Italy and are then available for deeper analysis.

Observations of \plnameB\ were performed on 2020 June 6 and June 19. Weather conditions were good with a windless clear sky, and air temperatures ranging from \SI{-71}{\degreeCelsius} to \SI{-69}{\degreeCelsius}  and from \SI{-77}{\degreeCelsius} to \SI{-73}{\degreeCelsius} on June 6 and June 19, respectively.

\begin{figure}
    \centering
    \includegraphics[width=0.8\hsize]{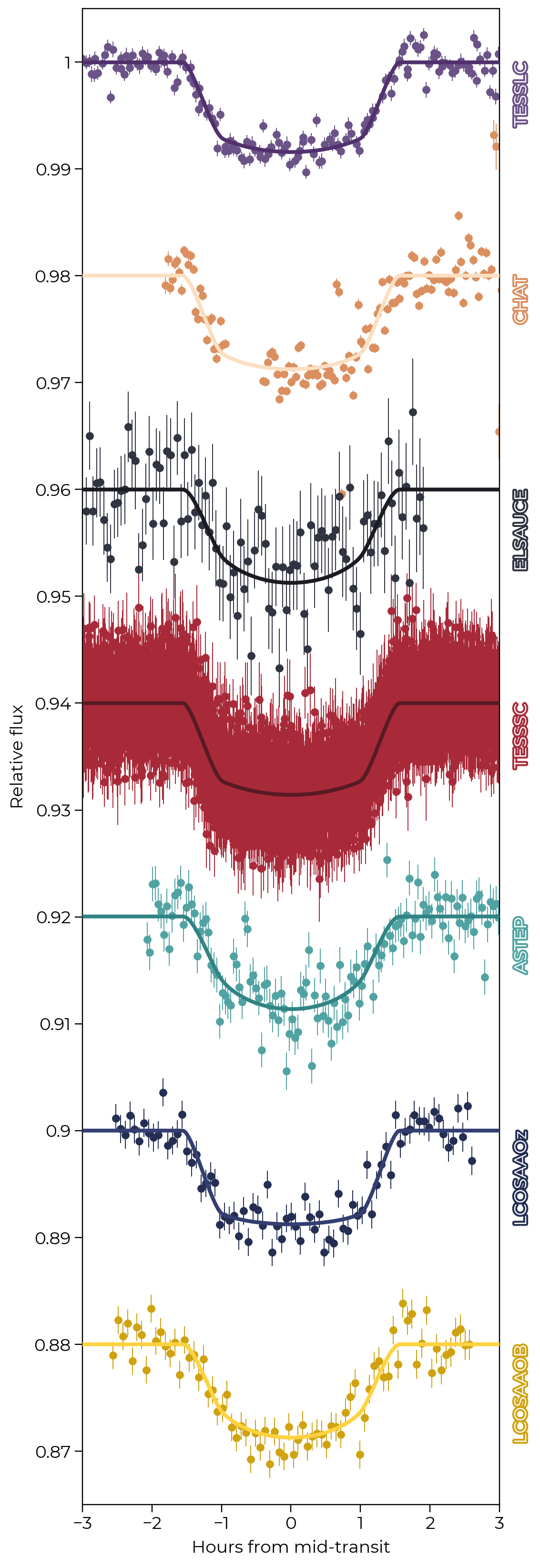}
    \caption{Transit data observed for \plnameB. The phase folded transit is plotted for both TESS long and short cadence. For CHAT and El Sauce, a single transit event is plotted.}
    \label{fig:phasedB}
\end{figure}

Full transits were observed on both 2020 June 6 and 19 observations with FWHMs of $6''$ and $5''$, respectively. The Moon, $\sim 99\%$ illuminated, was present during the 2020 June 6. The data were processed using IDL-based aperture photometry pipeline \citep{mekarnia2016}. The light curves, shown in \ref{fig:phasedB}, were obtained with a circular $11''$ photometric aperture.

\subsection{High resolution spectroscopy}

\subsubsection{FEROS}

Both stars were initially monitored with the FEROS spectrograph \citep{FEROS} mounted on the MPG 2.2m telescope at La Silla Observatory.

\stnameA\ was observed across 22 different epochs using FEROS from March 2019 to May 2019. An exposure time of 300 seconds was used, resulting in spectra with a signal-to-noise ratio (SNR) varying between 70 and 120, depending on the observing conditions and an average error on radial velocity of 6.65 m/s.

34 FEROS spectra of \stnameB\ were obtained between March 2019 and May 2019. We adopted an exposure time of 1200 seconds, which translated into spectra with a typical SNR of 55 and an average error of 10.51 m/s on the radial velocity.

In both cases, observations were performed with the simultaneous calibration mode for tracing the instrumental radial velocity drift during the science exposure. The FEROS data were processed with the \ceres\ pipeline \citep{ceres}. \ceres\ performs the optimal extraction of the spectra, wavelength calibration, instrumental drift correction, and barycentric correction and computes the radial velocity and bisector span measurements by using the cross-correlation technique. In this case, a binary mask resembling the spectral lines of a G2-type star was used as the cross-correlation template. The obtained radial velocities are presented in Tables \ref{tbl:rvA} and \ref{tbl:rvB}, for \stnameA\ and \stnameB, respectively. In both cases, the RVs were consistent with low amplitude (K$<$20 m s$^{-1}$) keplerian signal when adopting the TESS ephemeris, but the moderately large scatter around the best solution made evident the necessity of using more precise instruments to determine the mass of the planets.

\subsubsection{HARPS}
We used the HARPS spectrograph \citep{HARPS} mounted on the ESO 3.6m telescope at the ESO La Silla Observatory to further characterize the orbits of both transiting systems. 
For \stnameA , we obtained 14 HARPS spectra between April and December 2019. We used an exposure time of 900 s to reach a mean SNR of 80 and an average radial velocity error of 2.2 m/s. On the other hand, \stnameB\ was observed on 9 different epochs with HARPS between December 2019 and February 2020, adopting an exposure time of 1800 s for reaching a typical SNR of 40 and a average radial velocity error of 5.1 m/s.
The HARPS data was processed with the \ceres\ pipeline to obtain precision radial velocities and bisector span measurements which are listed in Tables \ref{tbl:rvA} and \ref{tbl:rvB}, and allowed us to confirm the planetary nature of both candidates and estimate the planetary masses.

\subsubsection{PFS}

\stnameA\ was monitored with the Planet Finder Spectrograph \citep[PFS,][]{pfs1,pfs2,pfs3} mounted on the Magellan/Clay telescope at Las Campanas Observatory. 10 observations were made using the iodine cell with an exposure time of 1500 seconds. Each observation was performed between April and October 2019. An additional 3$\times$1200 s iodine-free template was obtained in December of 2019 to compute the radial velocities as explained in \citet{butler:96}, obtaining an average radial velocity error of 0.92 m/s. PFS radial velocities for \stnameA\ are presented in Table \ref{tbl:rvA}.

\subsubsection{CHIRON}

For \plnameB, we acquired 6 epochs of observations with CHIRON spanning between November 2020 and January 2021, during the large program CARL-20B-3081 (PI: Carleo I.). CHIRON is a fiber fed high resolution echelle spectrograph on the SMARTS 1.5-meter telescope at Cerro Tololo Inter-American Observatory (CTIO) in Chile \citep{CHIRON}. We used the fiber slicer mode, which allows a resolution of $R\sim 80,000$ and a high efficiency \citep[for details on different available observing modes with CHIRON, see e.g.][]{CHIRON2}, and spreads the spectrum into 59 orders.

We split each visit into three individual exposures to achieve better cosmic ray rejection and acquired long-exposed ThAr lamp spectra at the start and end of each visit to monitor the wavelength drift of the instrument.  Since \plnameB\ is a relatively faint target, we used an exposure time of 1200 seconds, reaching an average radial velocity error of 32 m/s.

The spectra were extracted by making use of the official spectral extraction pipeline of CHIRON as per \citet{CHIRON2}.  The radial velocities were calculated by a least-squares deconvolution of the observation against a synthetic non-rotating ATLAS9 model atmosphere spectrum \citep{ATLAS9-2}. The least-squares deconvolution kernel is modeled via a broadening kernel encompassing the effects of radial velocity shift, rotational, instrumental, and macroturbulent broadening \citep{2021AJ....161....2Z}. In addition, the RV pipeline allows matching the observed spectrum against an interpolated grid of $\sim 10,000$ observed spectra pre-classified by the Spectroscopic Classification Pipeline \citep{2012Natur.486..375B} in order to estimate the spectral properties of the host star.

Radial velocities derived from CHIRON spectra are listed in Table \ref{tbl:rvB}.

\section{Analysis}
\label{sec:analysis}

\subsection{Stellar Parameters}
\label{sec:stpars}
We utilized the co-added FEROS spectra of each host star to determine their atmospheric parameters (\teff, \logg, \feh, \vsini) using the \zaspe\ code \citep{zaspe}. This code analyzes the observed high-resolution spectrum by comparing it to a grid of synthetic models created from the ATLAS9 model atmospheres \citep{castelli2004new}, focusing on spectral regions that are particularly sensitive to variations in the atmospheric parameters.

The physical parameters and evolutionary stages of the host stars are derived from the PARSEC \citep{parsec} stellar isochrones, along with the Gaia DR2 parallax, 2MASS \citep{2MASS}, Gaia broadband magnitudes and the atmospheric parameters obtained from \zaspe, as detailed in \citet{hd1397}. In summary, we identify the parameters of the PARSEC model that yield synthetic broadband photometric magnitudes that closely match the observed values while ensuring that the stellar effective temperature is compatible with that determined by \zaspe. We only consider PARSEC models with metallicities matching the spectroscopic values, and we convert the absolute magnitudes from these models into apparent magnitudes using the parallax and the extinction laws from \citet{cardelli}, considering the extinction coefficient A$_v$ as an additional free parameter. Through this method, we acquire the posterior parameters using the emcee package \citep{emcee}. This procedure yields the stellar mass and radius, allowing us to calculate a refined value for the stellar \logg\ that is more accurate than the spectroscopic estimate obtained via \zaspe. Given the correlations involved in determining the spectroscopic parameters, we repeat the \zaspe\ analysis again, fixing \logg\ to this new value, and then we repeat the determination of the stellar physical parameters, iterating until convergence in \logg\ is achieved.

The final atmospheric and physical parameters for \stnameA\ and \stnameB\ are listed in Table \ref{tab:stellarparams}. Both stars are G-type stars with very similar effective temperatures around 5700 K.

\begin{table*}
\caption{Stellar parameters of \stnameA\ and \stnameB.}
\label{tab:stellarparams}
\centering
\begin{threeparttable}

  \begin{tabular}{lccr}
   \hline
   \hline
     Parameter &  \stnameA\ &  \stnameB\ & Source\\
   \hline
Identifying Information & & & \\
~~~TIC ID & 123846039 &  177162886 & TIC$^a$\\
~~~GAIA ID & 3101010711776425088 & 5267242854094212224 &\textit{Gaia} DR2$^b$ \\
~~~2MASS ID & J06532863-0630270 & J06575823-7131230 & 2MASS$^c$\\
~~~R.A. (J2015.5, h:m:s) & $6^h53^m28.57^s$ & $6^h57^m58.19^s$ & \textit{Gaia} DR2$^b$ \\
~~~DEC (J2015.5, d:m:s) & $-6^\circ30'28.21''$ & $-70^\circ31'23.18''$ & \textit{Gaia} DR2$^b$\\
Proper motion and parallax & & & \\
~~~$\mu_\alpha \cos \delta$ (mas yr$^{-1}$) & -50.2 $\pm$ 0.1&  -6.95 $\pm$ 0.04 & \textit{Gaia} DR2$^b$ \\
~~~$\mu_\delta$ (mas yr$^{-1}$) & -76.0 $\pm$ 0.1 & 0.83 $\pm$ 0.04 & \textit{Gaia} DR2$^b$ \\
~~~Parallax (mas) & 9.71 $\pm$ 0.03 & 3.25 $\pm$ 0.02 & \textit{Gaia} DR2$^b$ \\
Spectroscopic properties  & & & \\
~~~$T_\textnormal{eff}$ (K) & \teffA & \teffB & ZASPE$^d$\\
~~~Spectral Type & G & G & ZASPE$^d$\\
~~~[Fe/H] (dex) & \fehA & \fehB & ZASPE$^d$\\
~~~$\log g_*$ (cgs)& \loggA & \loggB & ZASPE$^d$\\
~~~$v\sin(i_*)$ (km/s)& \vsiniA & \vsiniB & ZASPE$^d$\\
Photometric properties  & & & \\
~~~$T$ (mag)& $9.372 \pm 0.006$ & $11.587 \pm 0.006$ & TIC$^a$\\
~~~$G$ (mag)& $9.8279 \pm 0.0004$ & $12.0666 \pm 0.0002$ & \textit{Gaia} DR2$^b$ \\
~~~$B$ (mag)& $10.574\pm 0.071$ & $12.968\pm 0.027$ & Tycho-2$^e$\\
~~~$V$ (mag)& $9.957\pm0.005$ & $12.25 \pm0.010$ &Tycho-2$^e$\\
~~~$J$ (mag)& $8.749\pm0.026$ & $10.928\pm0.024$ & 2MASS$^c$\\
~~~$H$ (mag)& $8.437\pm0.038$ & $10.599\pm0.024$ & 2MASS$^c$\\
~~~$Ks$ (mag)& $8.37 \pm0.023$ & $10.51 \pm0.021$ & 2MASS$^c$\\
Derived properties  & & & \\
\vspace{0.1cm}
~~~$M_*$ ($M_\odot$)& \mstA & \mstB & PARSEC$^{*}$\\
~~~$R_*$ ($R_\odot$)& \rstA & \rstB & PARSEC$^{*}$\\
~~~$L_*$ ($L_\odot$)& \lstA & \lstB & PARSEC$^{*}$\\
~~~$A_v$ (mag) & $0.08^{+0.06}_{-0.05}$ & $0.18^{+0.069}_{-0.072}$ & PARSEC$^{*}$\\
~~~Age (Gyr)& \ageA & \ageB & PARSEC$^{*}$\\
~~~$\rho_*$ (g cm$^{-3}$)& \rhostA & \rhostB & PARSEC$^{*}$\\
   \hline
   \end{tabular}

\textit{Note}. Logarithms given in base 10.\\
(a) \textit{TESS} Input Catalog \citep{TIC8}; (b) \textit{Gaia} Data Release 2 \citep{gaiadr2summary}; (c) Two-micron All Sky Survey \citep{2MASS}; (d) Zonal Atmospheric Stellar Parameters Estimator \citep{brahm:2015,zaspe}; (e) Tycho-2 Catalog \citep{tycho2}\\
*: PARSEC isochrones \citep{parsec}; using stellar parameters obtained from ZASPE.
\end{threeparttable}
\end{table*}

\subsection{Global Modelling}
\label{sec:glob}

For both targets, we performed an analysis of the photometric and radial velocity data using \juliet\ \citep{juliet}, which allows us to do joint fits for photometry, using the \texttt{batman} \citep{batman} package, and RVs, using the \texttt{radvel} \citep{radvel} package. The parameter inference is done using nested sampling via the \texttt{MultiNest} package \citep{multinest} through \texttt{PyMultiNest} \citep{pymultinest} packages or the \texttt{dynesty} package \citep{dynesty}.

Nested sampling allows us to compute the log-evidences $\ln Z$ from the Bayesian model, which can be used for model comparison. When a $\Delta \ln Z > 3$ difference is present, the model with larger Bayesian log evidence is favored. If not, then both models are indistinguishable and the simpler model is chosen.

For the limb-darkening (LD) parametrization, we use a quadratic law for TESS light curves and a linear law for CHAT light curves. This selection was made following \cite{ChooseLD}. Both cases use the uninformative sampling scheme of \cite{KippingSampling}. Long integration times, such as TESS 30-min cadence, cause a smearing effect on light curves, biasing the parameters obtained with a model that assumes unbinned data. We follow the method of \cite{Kipping2010} and choose to resample long cadence light curves to have an integration time comparable to the short cadence ones. This implies resampling the 30-minute cadence light curves by a factor of 20.

Rather than using the planet-to-star radius ratio $p=R_p/R_{\star}$ and the impact parameter $b$, the parametrization $r_1$ and $r_2$ was used, previously defined in \citep{pbreparam}. These parameters ensure that all physically possible values in the (p,b) plane are explored.

Specific considerations for each system are described in section \ref{sec:joint}.

%Priors
\begin{table*}
    \caption{Adopted priors for the joint transit and RV analysis on both targets. $\mathcal{N}$, $\mathcal{U}$, $\mathcal{J}$ corresponds to normal, uniform and Jeffreys (log-uniform) distributions, respectively. Values reported are after choosing between eccentric and non-eccentric models.}
    \label{tbl:priors}
    \centering
        \begin{tabular}{lcccl} 
            \hline
            \hline
            Parameter name & Prior for \plnameA & Prior for \plnameB & Units & Description \\ \hline\hline\rule{0pt}{2ex}
            Star & & \\
            ~~~$\rho_*$ & $\mathcal{N}(1397,87^2)$ & $\mathcal{N}(1071,100^2)$ & kg/m$^3$ & Stellar density. \\
            \hline\rule{0pt}{2ex}
            Planet b\\
            ~~~$P_b$ & $\mathcal{N}(10.06,0.02^2)$ & $\mathcal{N}(12.84,0.02^2)$ & days & Period. \\
            ~~~$t_{0,b}$ & $\mathcal{N}(2458466.46,0.02^2)$ & $\mathcal{N}(2458313.63,0.02^2)$ & days & Time of transit-center. \\
            ~~~$r_{1,b}$ & $\mathcal{U}(0,1)$ & $\mathcal{U}(0,1)$ & --- & Parametrization for $p$ and $b$. \\
            ~~~$r_{2,b}$ & $\mathcal{U}(0,1)$ & $\mathcal{U}(0,1)$ & --- & Parametrization for $p$ and $b$. \\
            ~~~$K_{b}$ & $\mathcal{U}(0,100)$ & $\mathcal{U}(0,100)$ & m/s & Radial-velocity semi-amplitude. \\
            ~~~$e_b$ & 0 (fixed) & $\mathcal{U}(0,0.9)$ & --- & Eccentricity. \\
            ~~~$\omega$ & 90 (fixed) & $\mathcal{U}(0,360)$ & --- & Argument of periapsis. \\
            
            \hline\rule{0pt}{2ex}
            TESS (30-min)\\
            ~~~$D_{\textnormal{TESSLC}}$ & $\mathcal{U}(0,1)$ & $\mathcal{U}(0,1)$ & --- & Dilution factor for TESS. \\
            ~~~$M_{\textnormal{TESSLC}}$ & $\mathcal{N}(0,0.1^2)$ & $\mathcal{N}(0,0.1^2)$ & ppm & Relative flux offset for TESS. \\
            ~~~$\sigma_{w,\textnormal{TESSLC}}$ & $\mathcal{J}(10^{-4},5000)$ & $\mathcal{J}(0.1,5000)$ & ppm & Extra jitter term for \textit{TESS} lightcurve. \\
            ~~~$q_{1,\textnormal{TESSLC+TESSSC}}$ & $\mathcal{U}(0,1)$ & $\mathcal{U}(0,1)$ & --- & Quadratic LD parametrization. \\
            ~~~$q_{2,\textnormal{TESSLC+TESSSC}}$ & $\mathcal{U}(0,1)$ & $\mathcal{U}(0,1)$ & --- & Quadratic LD parametrization. \\

            \hline \rule{0pt}{2ex}
            TESS (2-min)\\
            ~~~$D_{\textnormal{TESSSC}}$ & 1 (fixed) & 1 (fixed) & --- & Dilution factor for TESS. \\
            ~~~$M_{\textnormal{TESSSC}}$ & $\mathcal{N}(0,0.1^2)$ & $\mathcal{N}(0,0.1^2)$ & ppm & Relative flux offset for TESS. \\
            ~~~$\sigma_{w,\textnormal{TESSSC}}$ & 0 (fixed) & 0 (fixed) & ppm & Extra jitter term for \textit{TESS} lightcurve. \\

            \hline\rule{0pt}{2ex}
            CHAT \\
            ~~~$D_{\textnormal{CHAT}}$ & 1 (fixed) & 1 (fixed) & --- & Dilution factor for CHAT. \\
            ~~~$M_{\textnormal{CHAT}}$ & $\mathcal{N}(0,0.1^2)$ & $\mathcal{N}(0,0.1^2)$ & ppm & Relative flux offset for CHAT. \\
            ~~~$\sigma_{w,\textnormal{CHAT}}$ & $\mathcal{J}(10^{-4},5000)$ & $\mathcal{J}(10^{-2},1000)$ & ppm & Extra jitter term for CHAT lightcurve. \\
            ~~~$q_{1,\textnormal{CHAT}}$ & $\mathcal{U}(0,1)$ & $\mathcal{U}(0,1)$ & --- & Linear LD parametrization. \\
            
            \hline\rule{0pt}{2ex}
            El Sauce \\
            ~~~$D_{\textnormal{ELSAUCE}}$ & --- & 1 (fixed) & --- & Dilution factor for El Sauce. \\
            ~~~$M_{\textnormal{ELSAUCE}}$ & --- & $\mathcal{N}(0,0.1^2)$ & ppm & Relative flux offset for El Sauce. \\
            ~~~$\sigma_{w,\textnormal{ELSAUCE}}$ & --- & $\mathcal{J}(10^{-2},1000)$ & ppm & Extra jitter term for El Sauce lightcurve. \\
            ~~~$q_{1,\textnormal{ELSAUCE}}$ & --- & $\mathcal{U}(0,1)$ & --- & Linear LD parametrization. \\    
            
            \hline\rule{0pt}{2ex}
            LCO (CTIO) \\
            ~~~$D_{\textnormal{LCOCTIO}}$ & 1 (fixed) & --- & --- & Dilution factor for LCO/CTIO. \\
            ~~~$M_{\textnormal{LCOCTIO}}$ & $\mathcal{N}(0,0.1^2)$ & --- & ppm & Relative flux offset for LCO/CTIO. \\
            ~~~$\sigma_{w,\textnormal{LCOCTIO}}$ & $\mathcal{J}(10^{-2},5000)$ & --- & ppm & Extra jitter term for LCO/CTIO. \\
            ~~~$q_{1,\textnormal{LCOCTIO}}$ & $\mathcal{U}(0,1)$ & --- & --- & Linear LD parametrization. \\ 
            
            \hline\rule{0pt}{2ex}
            LCO (MCD) \\
            ~~~$D_{\textnormal{LCOMCD}}$ & 1 (fixed) & --- & --- & Dilution factor for LCO/MCD. \\
            ~~~$M_{\textnormal{LCOMCD}}$ & $\mathcal{N}(0,0.1^2)$ & --- & ppm & Relative flux offset for LCO/MCD. \\
            ~~~$\sigma_{w,\textnormal{LCOMCD}}$ & $\mathcal{J}(10^{-2},5000)$ & --- & ppm & Extra jitter term for LCO/MCD. \\
            ~~~$q_{1,\textnormal{LCOMCD}}$ & $\mathcal{U}(0,1)$ & --- & --- & Linear LD parametrization. \\ 
            
            \hline \rule{0pt}{2ex}
            RV parameters\\
            ~~~$\mu_{\textnormal{FEROS}}$ & $\mathcal{N}(16617.1,20^2)$ & $\mathcal{N}(44695,20^2)$ & m/s & Systemic velocity for FEROS. \\
            ~~~$\sigma_{w,\textnormal{FEROS}}$ & $\mathcal{J}(10^{-6},0.2)$ & $\mathcal{J}(10^{-4},200)$ & m/s & Extra jitter term for FEROS. \\
            ~~~$\mu_{\textnormal{HARPS}}$ & $\mathcal{N}(16647.5,20^2)$ & $\mathcal{N}(44710,20^2)$ & m/s & Systemic velocity for HARPS. \\
            ~~~$\sigma_{w,\textnormal{HARPS}}$ & $\mathcal{J}(10^{-6},0.2)$ & $\mathcal{J}(10^{-4},200)$ & m/s & Extra jitter term for HARPS. \\
            ~~~$\mu_{\textnormal{PFS}}$ &$\mathcal{N}(0.002,0.2^2)$& --- & m/s & Systemic velocity for PFS. \\
            ~~~$\sigma_{w,\textnormal{PFS}}$ & $\mathcal{J}(10^{-6},0.2)$ & --- & m/s & Extra jitter term for PFS. \\
            ~~~$\mu_{\textnormal{CHIRON}}$ & --- & $\mathcal{U}(43000,44000)$ & m/s & Systemic velocity for CHIRON. \\
            ~~~$\sigma_{w,\textnormal{CHIRON}}$ & --- & $\mathcal{J}(10^{-4},200)$ & m/s & Extra jitter term for CHIRON. \\

            \hline
            \hline
        \end{tabular}
\end{table*}

\subsection{Radial Velocities}
We conducted a preliminary analysis using only radial velocities to identify additional non-transiting planets. We utilized the \texttt{RVSearch} pipeline \citep{rvsearch} to search iteratively for signals ranging from 1 day to the total observation baseline. Priors for period and time of conjunction for planet b, were established based on photometric data. The possibility of a linear trend was also explored. Only the transiting planets were recovered with sufficient significance for both targets. We independently validated this finding using an $l_1$ periodogram \citep{l1periodogram}, which analyzes all frequencies simultaneously, resulting in a cleaner power spectrum.

\begin{table*}
 \centering
 \caption{Evidences for preliminary RV analysis in \stnameA\ and \stnameB\ systems. No signs of non-transiting planets were found.}
 \label{tab:lnZRV}
\begin{tabular}{lccccc}
\hline
\hline
\multirow{2}{20em}{Model\hfill} & \multicolumn{2}{c}{\plnameA} & & \multicolumn{2}{c}{\plnameB} \\
& $\ln Z$ & $\Delta \ln Z$ & & $\ln Z$ & $\Delta \ln Z$ \\ \hline
1 planet circular                 & 147.85 & 0 & & 101.41 & 0 \\
1 planet eccentric                & 147.67 & -0.18 & & 99.72 & -1.69 \\
inner circular + outer circular   & 147.19 & -0.66 & & 98.17 & -3.24 \\
inner circular + outer eccentric  & 147.17 & -0.68 & & 98.45 & -2.96 \\
inner eccentric + outer circular  & 146.76 & -1.09 & & 96.51 & -4.9 \\
inner eccentric + outer eccentric & 145.95 & -1.9 &  & 96.16 & -5.25 \\
no planets                        & 121.09 & -26.76 & & 91.05 & -10.36 \\ \hline
\end{tabular}
\end{table*}

\subsection{Joint Analysis}
\label{sec:joint}
\begin{figure*}
    \centering
    \includegraphics[width=\hsize]{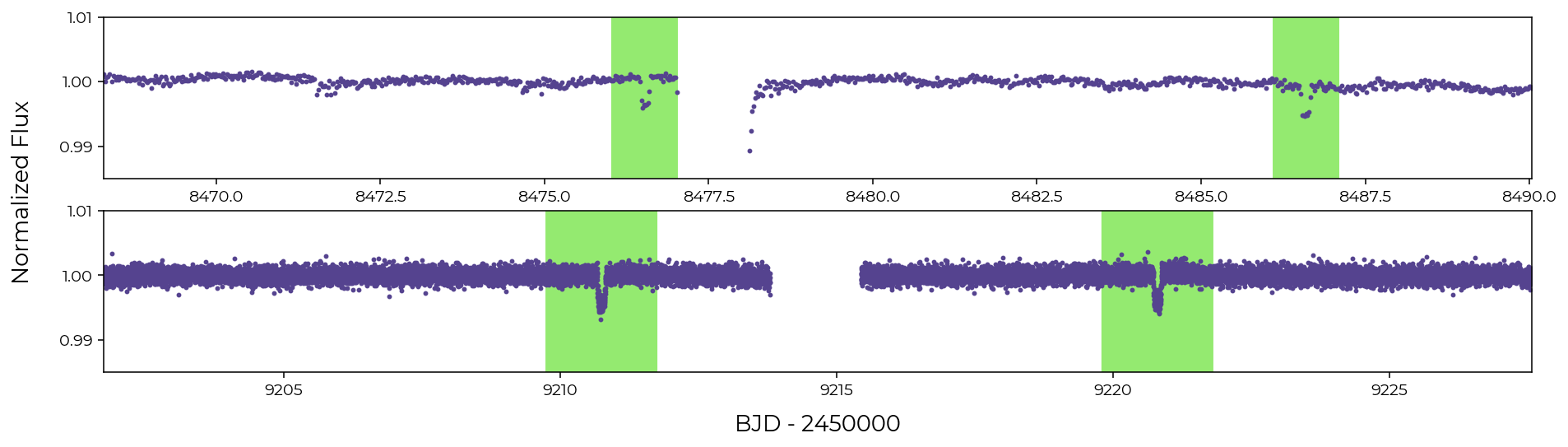}
    \caption{TESS Light curve for \stnameA. Top panel corresponds to long cadence data from Sector 6 and bottom panel short cadence from Sector 33.}
    \label{fig:TESSLCA}
\end{figure*}

\begin{figure*}
    \centering
    \includegraphics[width=\hsize]{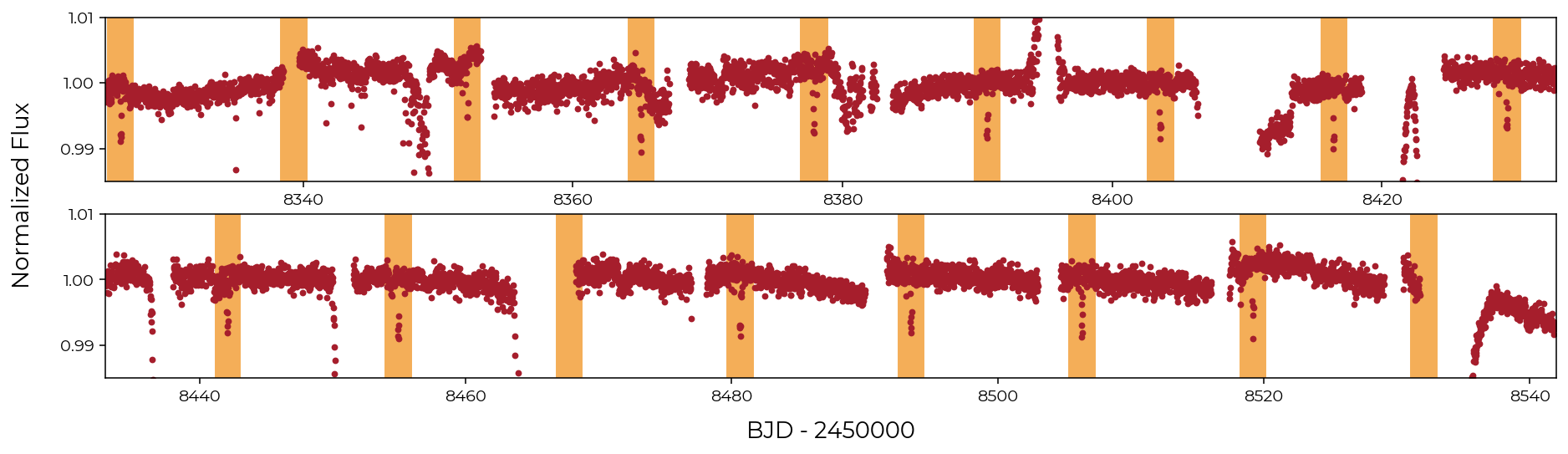}
    \caption{Long cadence TESS light curve for \stnameB. Data is taken from Sectors 1 to 8, with a total span of 217 days. \plnameB\  transits are highlighted in orange. Three of them are missing due to data gaps.}
    \label{fig:TESSLCB}
\end{figure*}

For the two systems, we performed a joint analysis of all available photometry and radial velocity data (TESS and ground-based) to infer the corresponding orbital parameters. For the photometric model, we included a Gaussian Process (GP) to account for underlying systematic or astrophysical signals present in the TESS long cadence data (see Figures \ref{fig:TESSLCA} and \ref{fig:TESSLCB}). As this adds significant computation time and we only care about the transits, we used only data within a 3-day window centered on each transit. A Matern 3/2 kernel was chosen for the GP model, which is parametrized by $\sigma_{GP}$ and $\rho$ as follows:

$$ k\left(\tau\right) = \sigma_{GP}^2\left(1+\frac{\sqrt{3}\tau}{\rho}\right) \exp\left(-\frac{\sqrt{3}\tau}{\rho}\right) $$

Using information previously obtained via BLS and RV-only analysis, we use normal priors for the period $P$, time of transit center $t_0$ and radial velocity offset $\mu$. The remaining parameters were explored using uniform or log-uniform priors, depending on whether the parameter space spanned several orders of magnitude. For non-eccentric models, eccentricity $e$ was fixed to $0$ and $\omega$ to $90$ degrees. A summary of all values used is listed in Table \ref{tbl:priors}.

\begin{figure*}
    \centering
    \begin{subfigure}{\hsize}
    \centering
    \includegraphics[width=0.24\hsize]{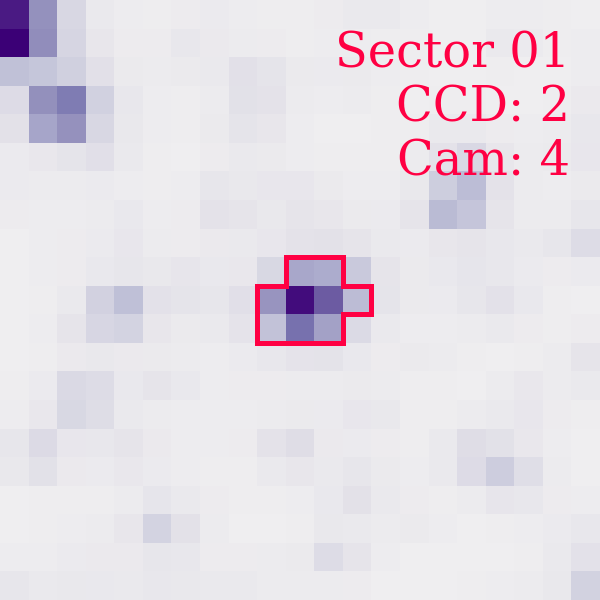}\hfill\includegraphics[width=0.24\hsize]{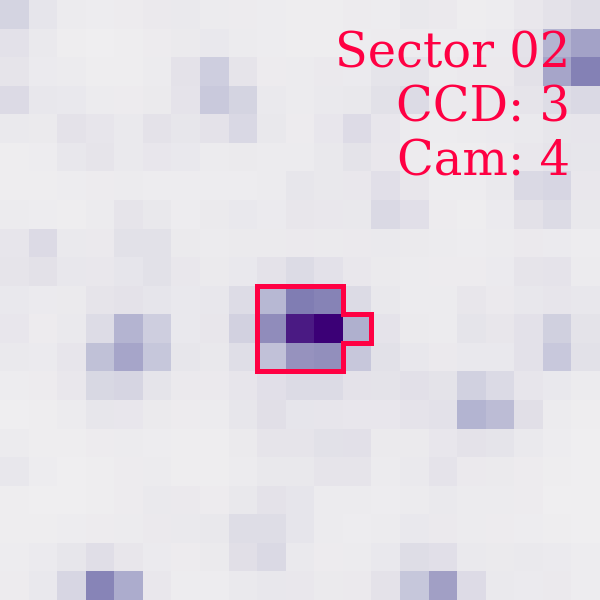}\hfill\includegraphics[width=0.24\hsize]{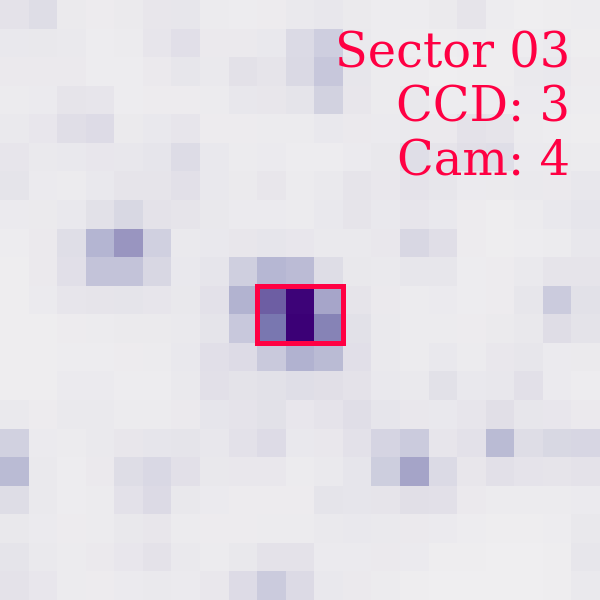}\hfill\includegraphics[width=0.24\hsize]{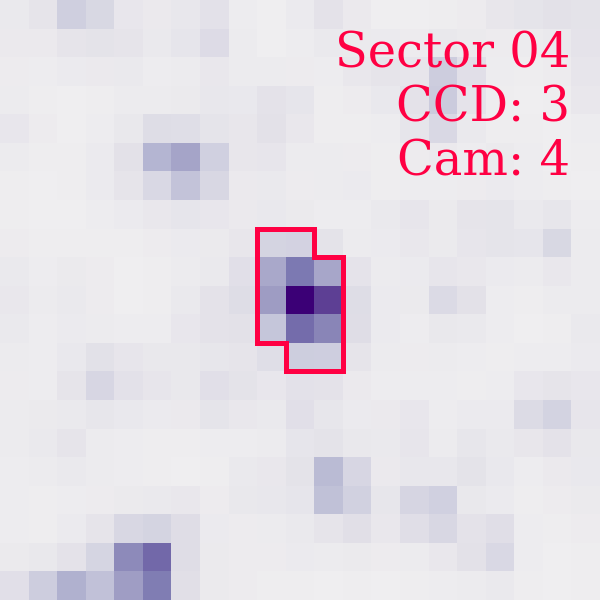}
    \end{subfigure}

   \vspace{.5em}

    \begin{subfigure}{\hsize}
    \centering
    \includegraphics[width=0.24\hsize]{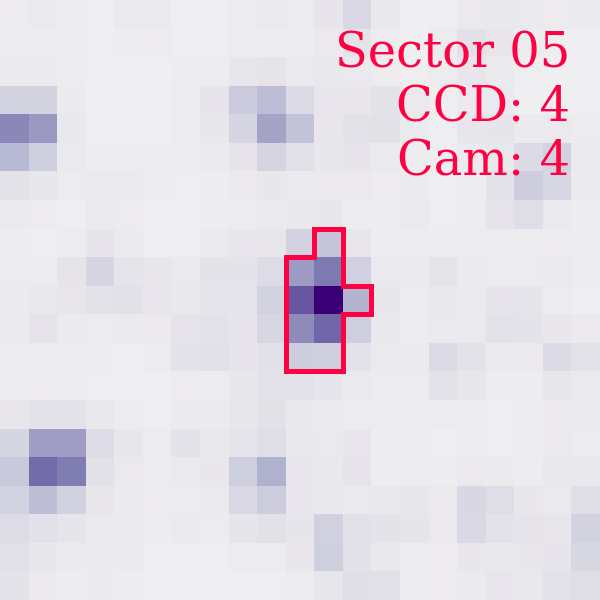}\hfill\includegraphics[width=0.24\hsize]{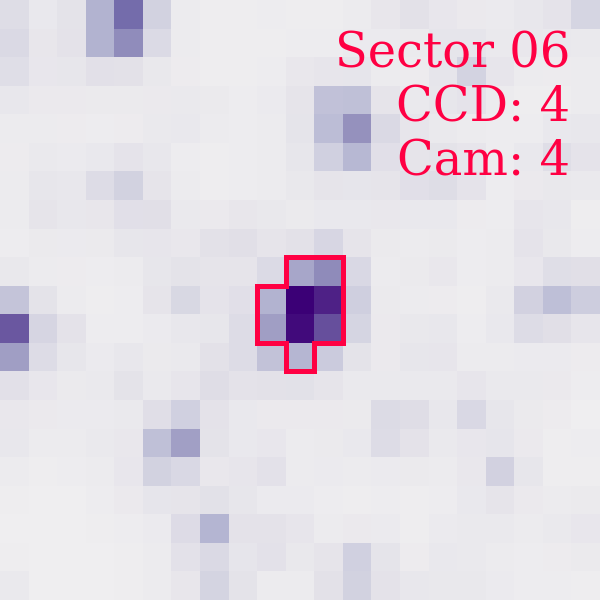}\hfill\includegraphics[width=0.24\hsize]{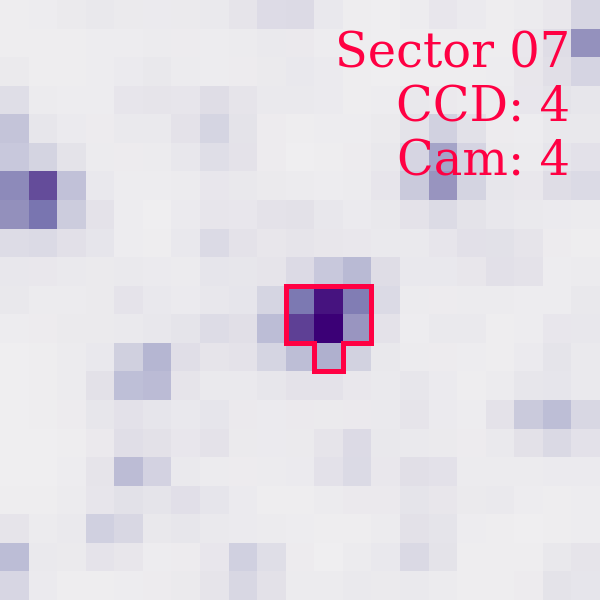}\hfill\includegraphics[width=0.24\hsize]{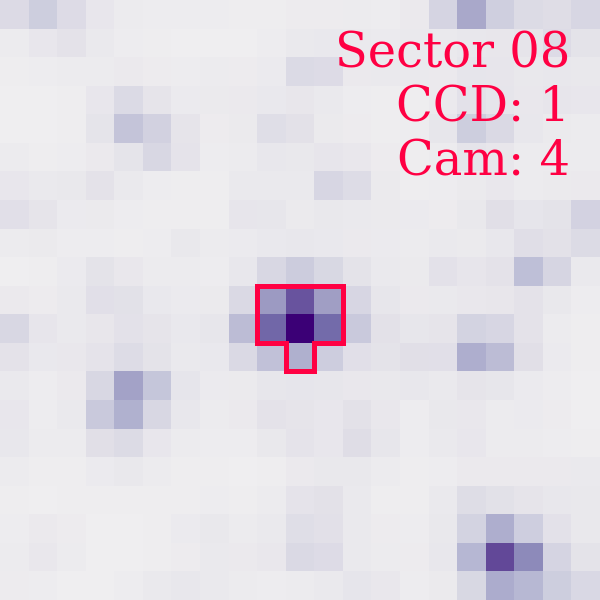}
    \end{subfigure}
    
    \caption{\stnameB\ as seen on TESS Sectors 1 to 8, which are only available in Long Cadence. As the field of view rotates, the star falls on different CCD, showing different shapes and neighborhoods, which can affect the noise of the light curve. The aperture chosen for each sector is highlighted in red.}
    \label{fig:stamps}
\end{figure*}

For \stnameB , each of the 23 sectors should ideally be considered as a different instrument. The reason behind this is that, for each sector, the target fell on different CCDs, cameras and pixels. In addition and as a consequence of the above, the pipeline makes different aperture choices (see Figure \ref{fig:stamps}). Considering all this, even with fixed apertures, each light curve should exhibit different noise and instrumental systematics. However, this implies a large increase in the number of parameters, making the analysis computationally intensive. To address this, a photometric fit of each sector was performed and compared the noise related parameters. Short and long cadence sectors were grouped and used as independent instruments instead.

Evidence for both models is reported in Table \ref{tbl:evidence}, which favors a non-eccentric model for \plnameA\ and eccentric model for \plnameB . Inferred parameters for this model are reported in Table \ref{tbl:posteriors}. The physical parameters of the systems are listed in Table \ref{tab:derivedparams}. We follow the suggestions from \cite{tayar2022}, adding their error floors to the stellar parameters in quadrature.

\begin{table}
    \caption{Calculated evidence returned by Juliet for the case comparing eccentric vs non-eccentric model. For \plnameA\ a non-eccentric model is favored, meanwhile, for \plnameB\ an eccentric one is chosen.}
    \label{tbl:evidence}
    \centering
    \begin{tabular}{lcc}
        \hline\hline
        %old ecc 181368.726 / e0 181362.457
        Model & \plnameA & \plnameB \\ \hline
        Eccentric & 112143.030 & 300242.699 \\
        Non-eccentric & 112158.073 & 300231.038 \\ \hline
    \end{tabular}
\end{table}

\begin{table*}
    \caption{\plnameA\ and \plnameB\ parameters.}
    \label{tbl:posteriors}
    \centering
        \begin{tabular}{lcccl} 
            \toprule
            \toprule
            Parameter name & Posterior for \plnameA & Posterior for \plnameB & Units & Description \\
            \midrule
            Parameters for the star & & \\
            ~~~$\rho_*$ & $1408^{+74}_{-78}$ & $1110^{+80}_{-86}$ & kg/m$^3$ & Stellar density. \\
            \midrule
            Parameters for planet b\\
            ~~~$P_b$ & $10.05772^{+0.00002}_{-0.00002}$ & $12.84618^{+0.00001}_{-0.00001}$ & days & Period. \\
            ~~~$t_{0,b}$ & $2458466.4730^{+0.0012}_{-0.0013}$ & $2458313.63758^{+0.00043}_{-0.00044}$ & days & Time of transit-center. \\
            ~~~$r_{1,b}$ & $0.657^{+0.023}_{-0.024}$ & $0.828^{+0.021}_{-0.023}$ & --- & Parametrization for $p$ and $b$. \\
            ~~~$r_{2,b}$ & $0.063^{+0.001}_{-0.001}$ & $0.094^{+0.001}_{-0.002}$ & --- & Parametrization for $p$ and $b$. \\
            ~~~$K_{b}$ & $11.926^{+0.961}_{-0.945}$ & $18.685^{+2.020}_{-1.990}$ & m/s & Radial-velocity semi-amplitude. \\
            ~~~$e_b$ & 0 (fixed) & $0.218^{+0.054}_{-0.057}$ & --- & Eccentricity. \\
            ~~~$\omega$ & 90 (fixed) & $128.526^{+14.743}_{-17.869}$ & --- & Argument of periapsis. \\
            
            \midrule
            Parameters for TESS (30-min)\\
            ~~~$D_{\textnormal{TESSLC}}$ & $0.972^{+0.018}_{-0.028}$ & $0.981^{+0.012}_{-0.016}$ & --- & Dilution factor for TESS. \\
            ~~~$\sigma_{w,\textnormal{TESSLC}}$ & $202^{+20}_{-20}$ & $739.800^{+25.445}_{-25.977}$ & ppm & Extra jitter term for \textit{TESS} lightcurve. \\
            ~~~$q_{1,\textnormal{TESSLC+TESSSC}}$ & $0.550^{+0.211}_{-0.185}$ & $0.253^{+0.075}_{-0.059}$ & --- & Quadratic LD parametrization. \\
            ~~~$q_{2,\textnormal{TESSLC+TESSSC}}$ & $0.123^{+0.148}_{-0.079}$ & $0.773^{+0.147}_{-0.209}$ & --- & Quadratic LD parametrization. \\

            \midrule
            Parameters for TESS (2-min)\\
            ~~~$D_{\textnormal{TESSSC}}$ & 1 (fixed) & 1 (fixed) & --- & Dilution factor for TESS. \\
            ~~~$\sigma_{w,\textnormal{TESSSC}}$ & 0 (fixed) & 0 (fixed) & ppm & Extra jitter term for \textit{TESS} lightcurve. \\

            \midrule
            Parameters for CHAT \\
            ~~~$D_{\textnormal{CHAT}}$ & 1 (fixed) & 1 (fixed) & --- & Dilution factor for CHAT. \\
            ~~~$\sigma_{w,\textnormal{CHAT}}$ & $3607^{+132}_{-121}$ & $997.838^{+1.491}_{-2.774}$ & ppm & Extra jitter term for CHAT lightcurve. \\
            ~~~$q_{1,\textnormal{CHAT}}$ & $0.495^{+0.237}_{-0.260}$ & $0.545^{+0.107}_{-0.119}$ & --- & Linear LD parametrization. \\
            
            \midrule
            Parameters for El Sauce \\
            ~~~$D_{\textnormal{ELSAUCE}}$ & --- & 1 (fixed) & --- & Dilution factor for El Sauce. \\
            ~~~$\sigma_{w,\textnormal{ELSAUCE}}$ & --- & $2.042^{+90.193}_{-1.976}$ & ppm & Extra jitter term for El Sauce lightcurve. \\
            ~~~$q_{1,\textnormal{ELSAUCE}}$ & --- & $0.785^{+0.138}_{-0.219}$ & --- & Linear LD parametrization. \\    
            
            \midrule
            Parameters for LCO (CTIO) \\
            ~~~$D_{\textnormal{LCOCTIO}}$ & 1 (fixed) & --- & --- & Dilution factor for LCO/CTIO. \\
            ~~~$\sigma_{w,\textnormal{LCOCTIO}}$ & $0.2^{+20}_{-0.2}$ & --- & ppm & Extra jitter term for LCO/CTIO. \\
            ~~~$q_{1,\textnormal{LCOCTIO}}$ & $0.39^{+0.16}_{-0.17}$ & --- & --- & Linear LD parametrization. \\ 
            
            \midrule
            Parameters for LCO (MCD) \\
            ~~~$D_{\textnormal{LCOMCD}}$ & 1 (fixed) & --- & --- & Dilution factor for LCO/MCD. \\
            ~~~$\sigma_{w,\textnormal{LCOMCD}}$ & $0.17^{+23.66}_{-0.16}$ & --- & ppm & Extra jitter term for LCO/MCD. \\
            ~~~$q_{1,\textnormal{LCOMCD}}$ & $0.710^{+0.179}_{-0.225}$ & --- & --- & Linear LD parametrization. \\ 
            
            \midrule
            RV parameters\\
            ~~~$\mu_{\textnormal{FEROS}}$ & $16615.36^{+2.68}_{-2.57}$ & $44697.329^{+3.834}_{-3.926}$ & m/s & Systemic velocity for FEROS. \\
            ~~~$\sigma_{w,\textnormal{FEROS}}$ & $10.51^{+2.50}_{-2.07}$ & $23.098^{+3.128}_{-2.750}$ & m/s & Extra jitter term for FEROS. \\
            ~~~$\mu_{\textnormal{HARPS}}$ & $16648.28^{+1.39}_{-1.34}$ & $44713.701^{+1.848}_{-1.709}$ & m/s & Systemic velocity for HARPS. \\
            ~~~$\sigma_{w,\textnormal{HARPS}}$ & $4.65^{+1.34}_{-0.99}$ & $0.021^{+0.572}_{-0.020}$ & m/s & Extra jitter term for HARPS. \\
            ~~~$\mu_{\textnormal{PFS}}$ & $0.82^{+0.86}_{-0.88}$ & --- & m/s & Systemic velocity for PFS. \\
            ~~~$\sigma_{w,\textnormal{PFS}}$ & $2.64^{+0.88}_{-0.62}$ & --- & m/s & Extra jitter term for PFS. \\
            ~~~$\mu_{\textnormal{CHIRON}}$ & --- & $43302.230^{+10.865}_{-11.197}$ & m/s & Systemic velocity for CHIRON. \\
            ~~~$\sigma_{w,\textnormal{CHIRON}}$ & --- & $0.068^{+2.092}_{-0.067}$ & m/s & Extra jitter term for CHIRON. \\

            \bottomrule
        \end{tabular}
\end{table*}

\subsection{Transit Timing Variations of \plnameB}
\begin{figure*}
    \centering
    \includegraphics[width=\hsize]{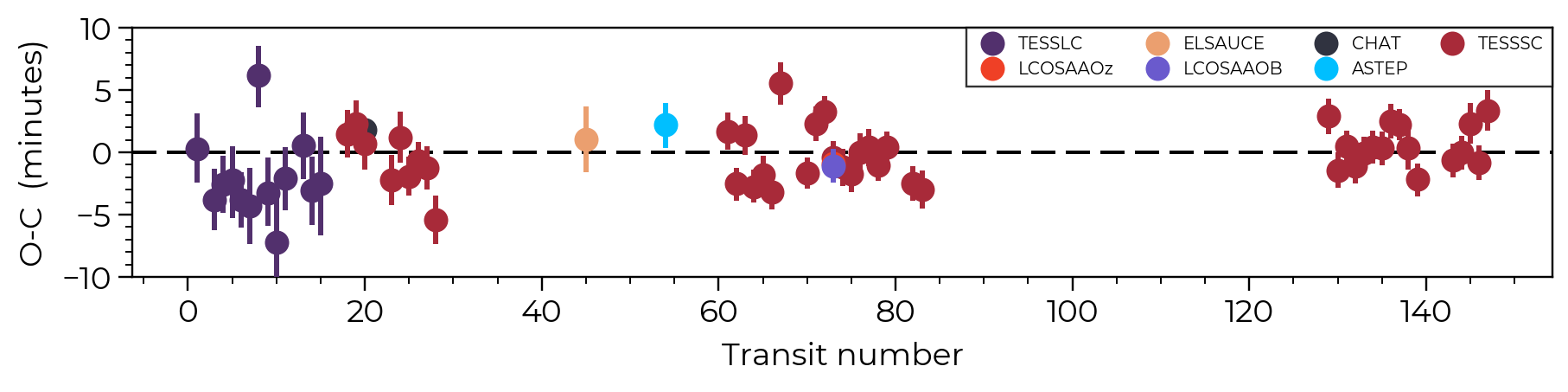}
    \caption{Observed minus calculated diagram of all the transits observed for \plnameB. No significant periodic signal was found in the data, ruling out an additional companion found via TTVs.}
    \label{fig:ttvs}
\end{figure*}

Given that \stnameB\ was observed in the 13 TESS sectors of the southern hemisphere, we searched for transit timing variations of \plnameB\ that could be induced by additional planets in the system. We ran an independent \juliet\ fit on the photometry, leaving the time of transit as a free parameter and using all the transit data available: 13 from TESS in long cadence, 26 from TESS in short cadence, plus 2 from the follow-up data acquired with CHAT and El Sauce. We adopted normal priors using the results from Table \ref{tbl:posteriors}. The difference between observed and predicted transit time for \plnameB\ is presented in Figure \ref{fig:ttvs} as a function of the transit number. No significant variations in transit timing are identified.

\begin{figure*}
    \centering
    \includegraphics[width=\hsize]{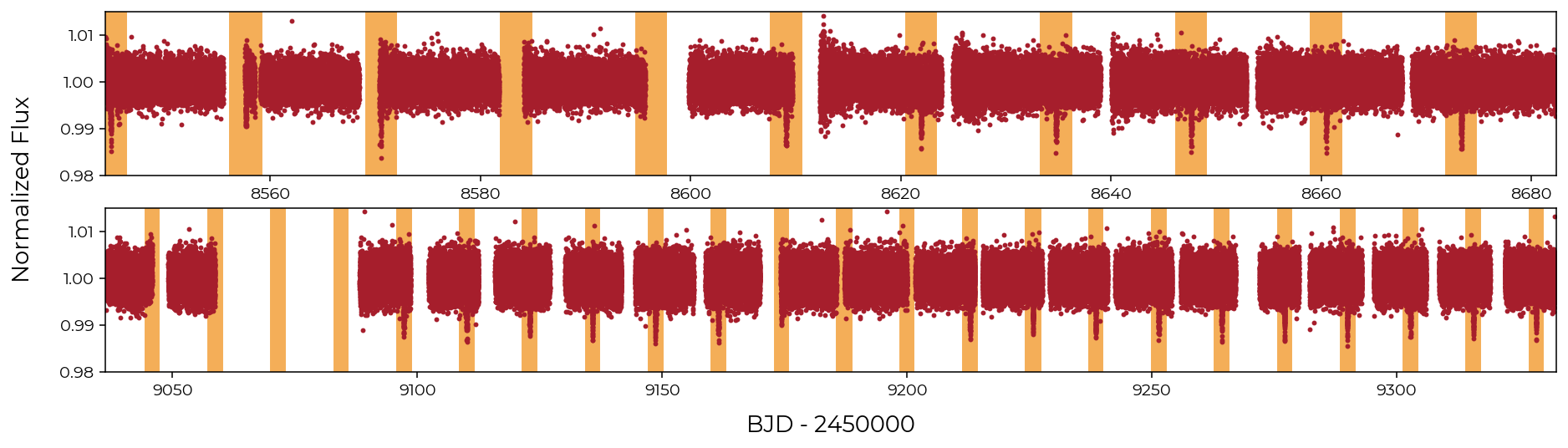}
    \caption{Short cadence TESS light curve for \stnameB. Top panel shows data from Sectors 9 to 13. Bottom data comes from Sectors 27 and 29 to 37. Transit occurrences are highlighted in orange.}
    \label{fig:TESSSCB}
\end{figure*}

\begin{figure*}
    \centering
    \includegraphics[width=\hsize]{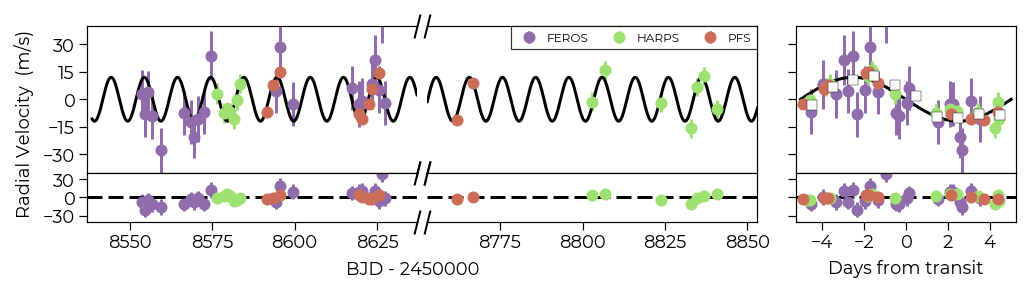}
    \caption{Radial velocities observed for \stnameA\ with FEROS (purple), HARPS (green) and PFS (orange). Phase folded radial velocity curve is plotted in the right panel. The best Keplerian model is included and plotted in black.}
    \label{fig:rvsA}
\end{figure*}

\begin{figure*}
    \centering
    \includegraphics[width=\hsize]{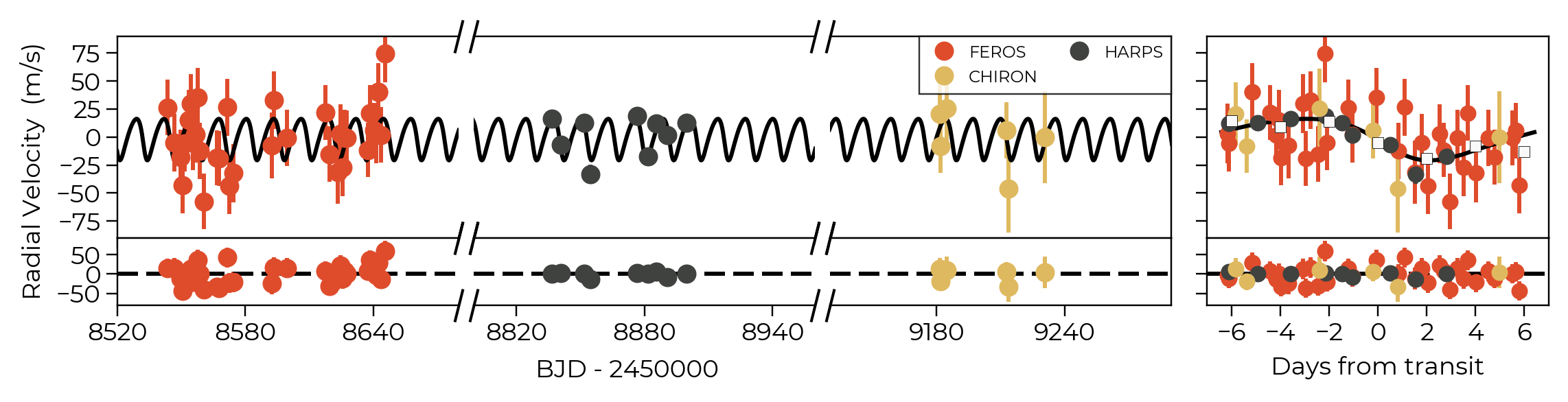}
    \caption{Radial velocities from FEROS (orange-red), HARPS (black) and CHIRON (yellow) observed for \stnameB. The best Keplerian model is plotted as a black curve in both panels.}
    \label{fig:rvsB}
\end{figure*}

\begin{figure*}
    \centering
    \includegraphics[width=0.5\hsize]{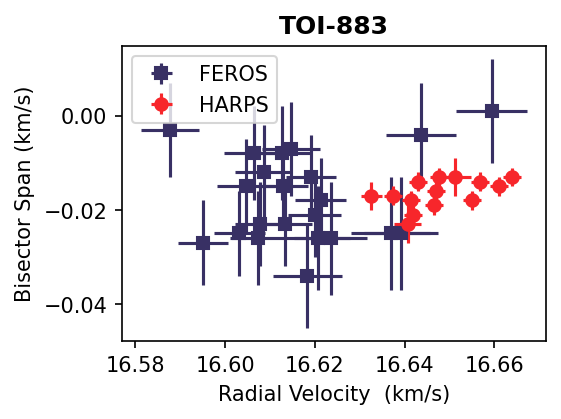}\includegraphics[width=0.5\hsize]{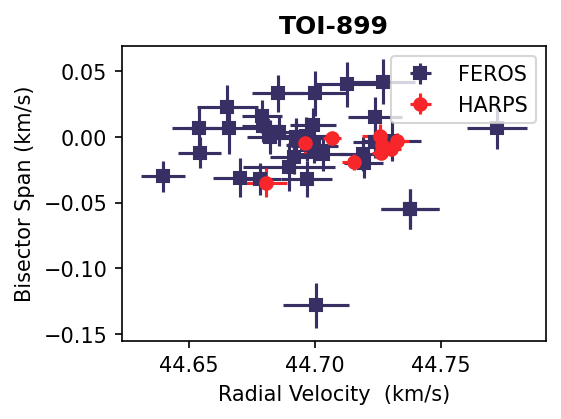}
    \caption{Radial velocity data against the corresponding bisector span measured for each observation. No correlation was found between both targets.}
    \label{fig:bisector}
\end{figure*}

%% Table of posterior values (Planetary Parameters)
\begin{table*}
    \centering
    \caption{Derived planetary parameters obtained for \plnameA\ and \plnameB\ using the posterior values from Table \ref{tbl:posteriors}. $T_{eq}$ was calculated using equation 16 from \cite{teq}, assuming an albedo $A=0$ and a heat redistribution $\beta = 0.5$}
    \label{tab:derivedparams}
    \begin{tabular}{lcc} 
        \toprule
        \toprule
        \multirow{2}{10em}{Parameter name} & \multicolumn{2}{c}{Posterior estimate} \\
        & \plnameA & \plnameB \\
        \midrule
        Posterior parameters & & \\
        ~~~$P_b$ (days) &$10.057716^{+0.000016}_{-0.000016}$ & $12.846185^{+0.000008}_{-0.000008}$\\[0.1 cm]
        ~~~$t_{0,b}$ (BJD UTC) &$2458466.473005^{+0.001214}_{-0.001252}$ & $2458313.637577^{+0.000434}_{-0.000441}$ \\[0.1 cm]
        ~~~$\rho_*$ (kg/m$^3$) & $1408^{+74}_{-78}$ & $1110^{+80}_{-86}$\\[0.1 cm]
        ~~~$r_{1,b}$ & $0.6572^{+0.0232}_{-0.0240}$ & $0.8277^{+0.0206}_{-0.0229}$\\[0.1 cm]
        ~~~$r_{2,b}$ & $0.0628^{+0.0008}_{-0.0008}$ & $0.0937^{+0.0012}_{-0.0015}$\\[0.1 cm]
        ~~~$K_{b}$ (m/s) & $11.926^{+0.961}_{-0.945}$ & $18.685^{+2.020}_{-1.990}$\\[0.1 cm]
        ~~~$e_b$ & $0$ (fixed) & $0.218^{+0.054}_{-0.057}$\\[0.1 cm]
        \midrule
        Transit parameters  & & \\
        ~~~$R_p/R_{\star}$ & $0.0628^{+0.0008}_{-0.0008}$ & $0.0937^{+0.0012}_{-0.0015}$\\[0.1 cm]
        ~~~$b = (a/R_{\star})\cos(i_p)$ & $0.49^{+0.03}_{-0.04}$ & $0.742^{+0.030}_{-0.034}$ \\[0.1 cm]
        ~~~$a_{b}/R_{\star}$ & $19.55^{+0.99}_{-0.99}$ & $21.316^{+0.502}_{-0.564}$\\[0.1 cm]
        ~~~$i_p$ (deg) & $88.58^{+0.12}_{-0.13}$ & $87.572^{+0.125}_{-0.150}$\\[0.1 cm]
        ~~~$q_{1,TESS}$  & $0.55^{+0.21}_{-0.18}$ & $0.25^{+0.07}_{-0.06}$\\[0.1 cm]
        ~~~$q_{2,TESS}$  & $0.12^{+0.15}_{-0.08}$ & $0.77^{+0.15}_{-0.21}$ \\[0.1 cm]
        \midrule
        Physical parameters & &\\[0.1cm]
        ~~~$M_{p}$ ($M_J$) & $0.123\pm 0.012$ & $0.213\pm 0.024$\\[0.1 cm]
        ~~~$R_{p}$ ($R_J$) & $0.604\pm 0.028$ & $0.991\pm 0.044$\\[0.1 cm]
        ~~~$a$ (AU) & $0.0898\pm 0.0023$ & $0.1063\pm 0.0026$\\[0.1 cm]
        ~~~$T_\textnormal{eq}$ (K) & $1086\pm 19$ & $1040\pm 19$ \\[0.1 cm]
        \bottomrule
    \end{tabular}
\end{table*}

\section{Discussion}
\label{sec:discussion}

\begin{figure}
    \centering
    \includegraphics[width=\hsize]{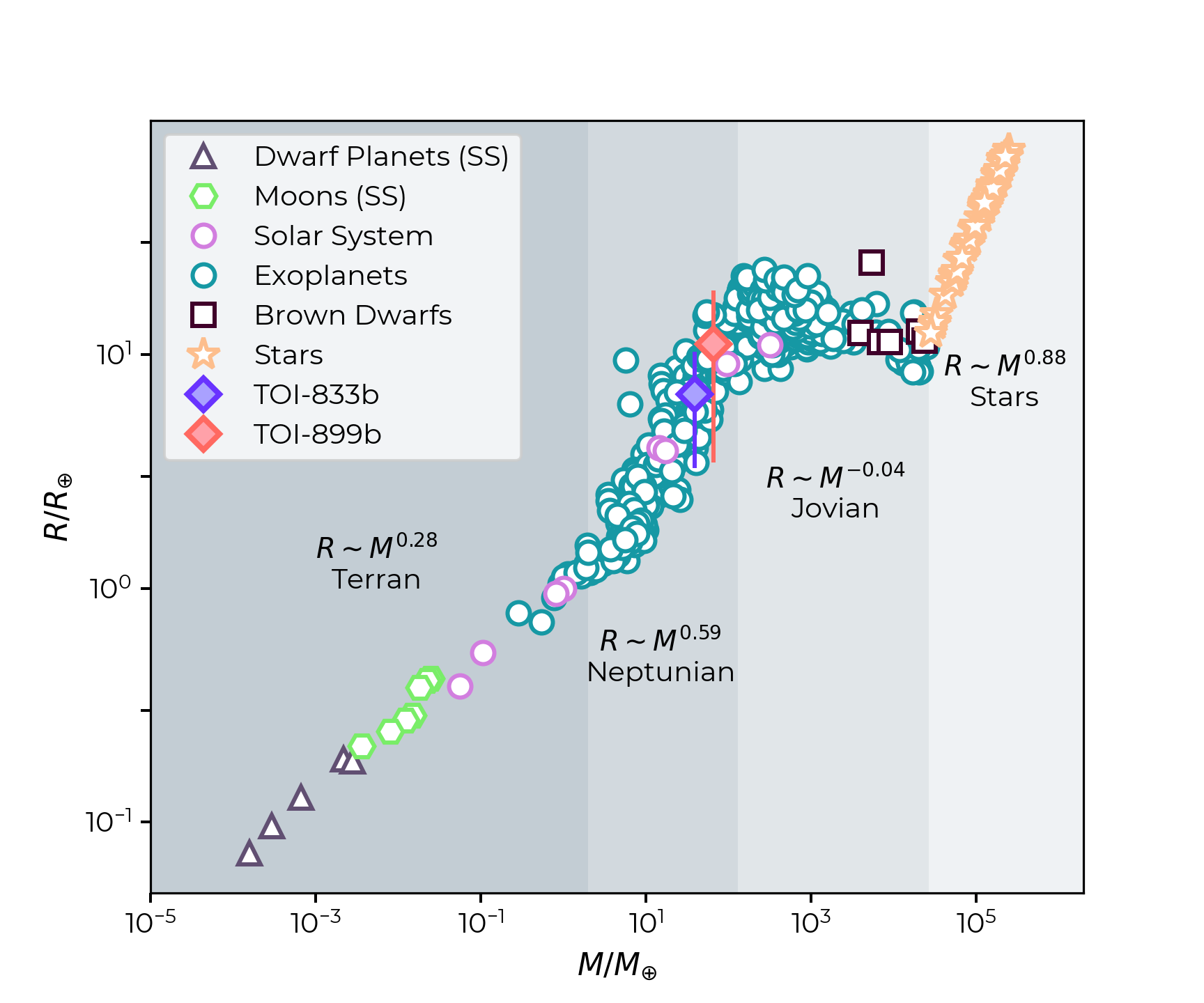}
    \caption{Mass vs Radius diagram featuring the classification developed by \protect\cite{Chen2017}. Both targets are located close to the transition between Neptunian and Jovian planets. Exoplanet data are taken from the TEPCat catalog \citep{TEPCAT}.}
    \label{fig:RvsP}
\end{figure}

Figure \ref{fig:RvsP} shows how the physical properties of the two new transiting planets presented in this study compare to those of the current population of planetary and sub-stellar objects. According to the classification presented in \citet{Chen2017} \plnameA\ and \plnameB\ fall in the so-called Neptunian region, where the steepest variation in radius as a function of mass is observed for planetary objects. This abrupt change in their physical properties is possibly produced by the different envelope accretion histories that these objects can present before reaching a limiting mass around that of Saturn, where a gaseous dominated structure becomes partially electron degenerate.

\begin{figure*}
    \centering
    \includegraphics[width=\hsize]{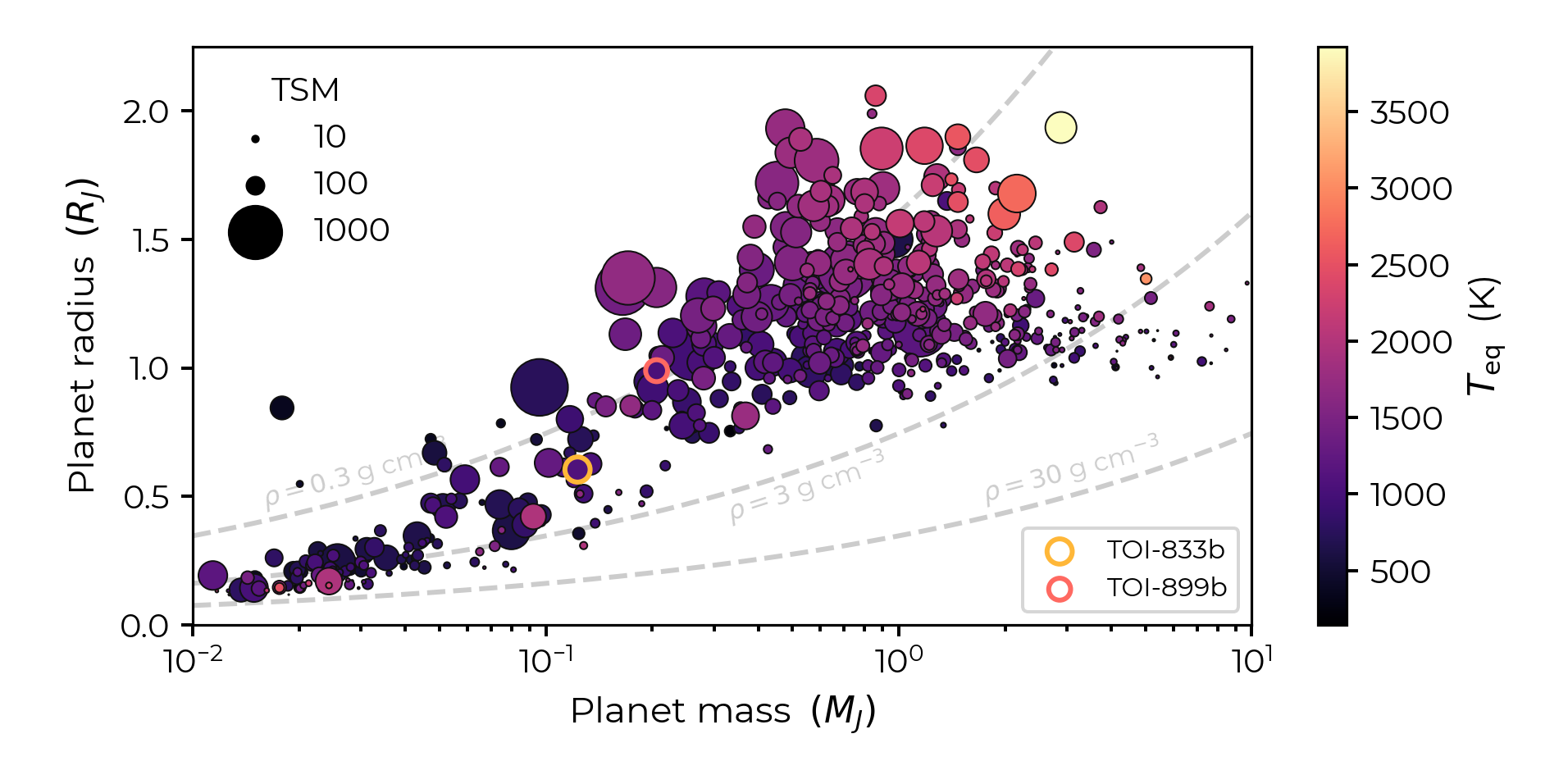}
    \caption{Mass vs Radius diagram of well-characterized transiting planets \protect\citep{TEPCAT}. \plnameA\ and \plnameB\ are highlighted in yellow and red, respectively. Dot size scales with the Transmission Spectroscopy Metric (TSM) defined by \protect\cite{TSM}.}
    \label{fig:tepcat}
\end{figure*}

\plnameA\ and \plnameB\ orbit main sequence solar-type stars with solar-like metallicities and have similar orbital periods (10.1 days and 12.8 days, respectively). Despite this, the two planets seem to have remarkably different physical structures, which is further highlighted in Figure \ref{fig:tepcat}. While the radius of \plnameA\ is relatively similar to that of Neptune (1.76 Neptune radii), \plnameB\ has an inflated radius of 2.88 Neptune radii. Their masses also show some discrepancy, with \plnameA\ having a mass of 0.12 Jupiter masses and \plnameB\ 0.2 Jupiter masses. Due to their moderately long periods and mild insolation levels, the internal structure of both planets is not expected to be significantly affected by stellar radiative, tidal, and/or magnetic effects \citep{spiegel:2013}, and in turn, these differences in structure can inform us about the varied composition that a giant planet can have depending on the disk conditions where they were formed.

The radius of \plnameA\ aligns with traditional structural models by suggesting a moderate enrichment of heavy elements. Utilizing the models from \citet{fortney:2007}, which assume that all heavy elements are situated in the planet's core, we determine that a core mass of $M_{c}=25.03 \pm 2.55$ \mearth\ is adequate to account for the planet’s mass and radius. A slightly lower level of heavy element enrichment could be achieved with more recent models \citep[e.g.][]{thorngren:2016}, which allows some solids to be distributed within the gaseous envelope. \plnameA , according to some recent formation models, could have formed at long distances from the central star, where high envelope luminosity due to the constant inflow of pebbles prevents the core from reaching the threshold mass required to trigger the runaway gaseous accretion before the disk dispersal \citep{bitsch:2015,chen:2020,brouwers:2021}. The challenge in adopting this formation scenario for \plnameA\ is to explain its current short semi-major axis. A disk-driven migration for this planet would suggest accessing the inner areas of the disk where the pebble isolation mass is reachable, leading to the rapid development of a gas giant. Scenarios involving high eccentricity migration \citep[e.g.][]{rasio:1996,wu:2003,naoz:2012}, particularly after the disk has dissipated, are plausible explanations for the current characteristics of \plnameA . This could also account for its mild eccentricity and the absence of other large, nearby planets in the system. Among the currently known population of well-characterized transiting planets, \plnameA\ closely resembles HATS-7b \citep{bakos:2015}, which has a radius of 0.57 \rjup\ and a mass of 0.12 \mjup\. In comparison, \plnameA\ has a radius of 0.605 \rjup\ and a mass of 0.124 \mjup\. However, HATS-7b orbits a cooler K-type star and has a shorter orbital period of 3.2 days. Additionally, K2-234b \citep{yu:2018} shares similar physical characteristics with \plnameA\ and has a moderately longer orbital period of 11.8 days, but it orbits a slightly more massive subgiant star.

\plnameB, on the other hand, is a significantly less dense sub-Saturn.
By using the \citet{fortney:2007} models for the properties of the \stnameB\ system, we find that this planet is consistent with having a core-less interior (M$_{c}$=0.29 $\pm$ 1.53 \mearth), being completely dominated by a gaseous structure. Specifically, we can put an upper limit of 4 \mearth\ for the bulk heavy element enrichment of \plnameB. The other three sub-Saturns with low irradiation and masses compatible with \plnameB\ \citetext{\citealp[Kepler-34b,][]{welsh:2012}; \citealp[TOI-1456b,][]{dalba:2020}; \citealp[TOI-197b,][]{huber:2019}}, all have radii below 0.9 \rjup\ which implies the presence of a massive central core of 10-20 \mearth, more in line with the expectations of the core accretion theory. 

Both planets presented here pose an additional challenge to planet formation models: their masses fall into a valley of low occurrence predicted by core accretion models~\citep[e.g.,][]{Ida2004,Mordasini2009b,Emsenhuber2021b,Burn2021,Schlecker2021b,Schlecker2022} due to the rapid growth via runaway gas accretion of planets reaching a mass above $\sim 10 \mearth$~\citep{Mizuno1978,pollack:96}.
This predicted bimodality in the mass distribution of giant exoplanets has been tentatively detected in homogeneously derived samples of radial velocity-detected planets~\citep{Mayor2011} but appears absent in more recently collected planet populations discovered with the RV~\citep{Sabotta2021,Schlecker2022,Ribas2023} and microlensing techniques~\citep{Suzuki2016,Suzuki2018,Bennett2021}.
The absence of this ``sub-Saturn valley'' has been hypothesized to point to oversimplifications in classical 1D hydrostatic gas accretion models~\citep{Szulagyi2014,Moldenhauer2021,Schlecker2022,Krapp2022}.
Planets on intermediate orbits and with masses like \plnameA\ and \plnameB\ are a valuable contribution to a sparse sample that will help to characterize the existence and shape of the mass valley, which may constrain planetary gas accretion theories.

\begin{figure*}
    \centering
    \includegraphics{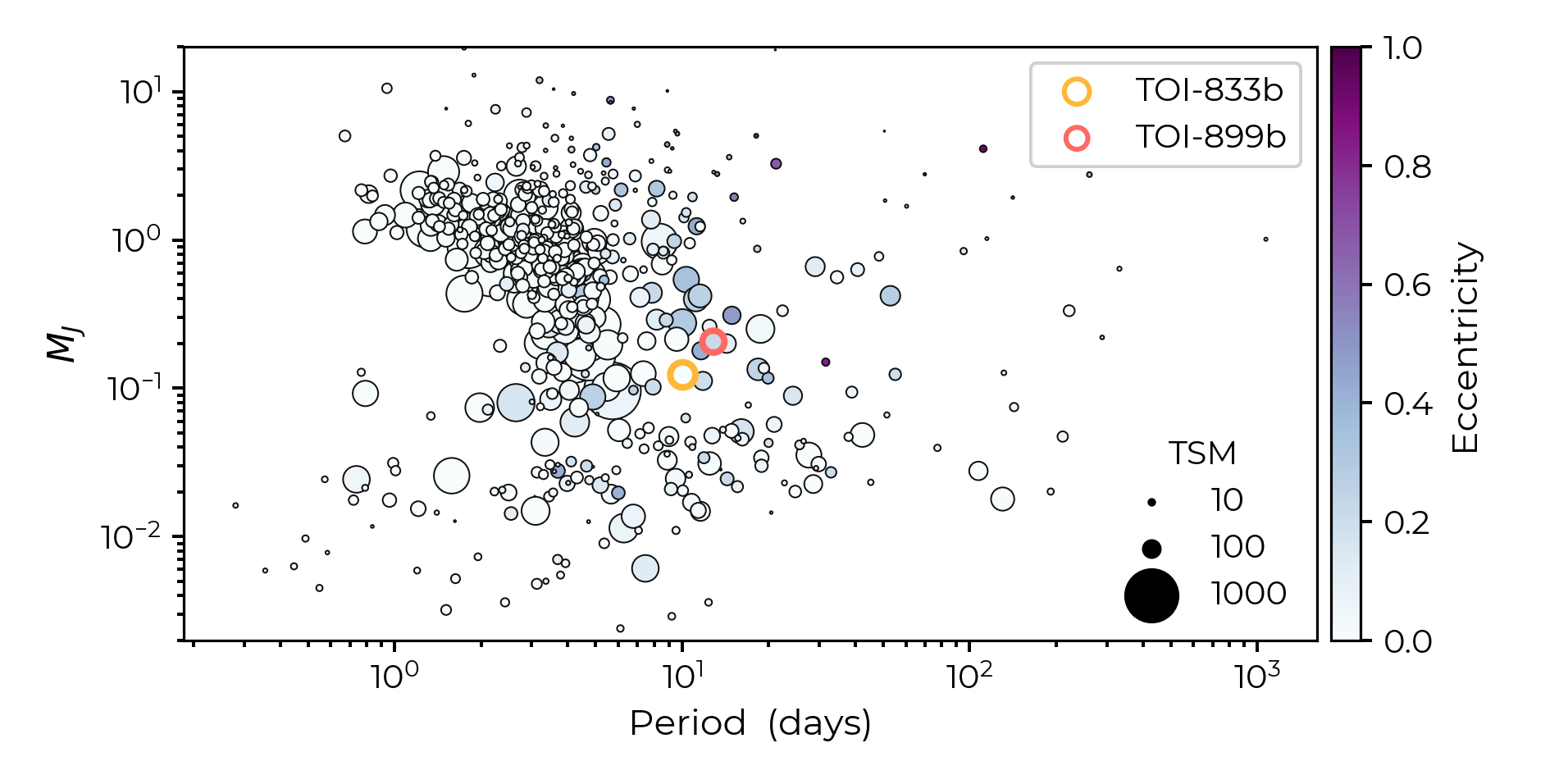}
    \caption{Mass vs period diagram. This includes the TSM metric to compare against other similar Warm Giants. Eccentricity is also included, showing a noticeable contrast between the Warm Jupiter (some eccentric) and Hot Jupiter (mostly circular) populations.}
    \label{fig:TSMe}
\end{figure*}

Future detailed follow-up observations of \plnameA\ and \plnameB\ could contribute to unveiling their formation histories in this and other regards. 
Having transmission spectroscopy metrics \citep{kempton:2018} of 235 and 180, respectively, both systems are suitable targets for atmospheric characterization (see Figure \ref{fig:TSMe}). In this context, the determination of their atmospheric metallicities \citep[e.g.][]{wong:2020} could further help to constrain the properties of the regions of the protoplanetary disk where they were formed. Additionally, due to their moderately long periods and mild tidal interactions, both systems are interesting targets for measuring their projected stellar obliquities through the Rossiter-McLaughlin (RM) effect for constraining migration scenarios. \plnameA\ and \plnameB\ have expected RM semi-amplitude of 5.8 ms$^{-1}$ m/s and 14 ms$^{-1}$, respectively for aligned orbits.

%RVs for target A
\begin{table}
    \centering
    \caption{Radial velocity data obtained for \plnameA .}
    \label{tbl:rvA}
    \centering
    \begin{tabular}{lcc}
    \toprule
    \toprule
    BJD - 2450000 & Radial Velocity  (m/s) & Bisector (m/s) \\
    \midrule
    FEROS & & \\
    ~~~8553.67181 & $16618.3 \pm 77$ & $-34 \pm 11$ \\
    ~~~8627.45229 & $16613.4 \pm 60$ & $-23 \pm 9$ \\
    ~~~8626.45925 & $16659.4 \pm 79$ & $1 \pm 11$ \\
    ~~~8625.45179 & $16620.7 \pm 72$ & $-26 \pm 11$ \\
    ~~~8624.45751 & $16636.9 \pm 88$ & $-25 \pm 12$ \\
    ~~~8623.44687 & $16623.5 \pm 81$ & $-26 \pm 12$ \\
    ~~~8620.45369 & $16614.7 \pm 65$ & $-7 \pm 10$ \\
    ~~~8619.46761 & $16612.8 \pm 64$ & $-8 \pm 10$ \\
    ~~~8617.46982 & $16621.3 \pm 57$ & $-18 \pm 9$ \\
    ~~~8599.49535 & $16612.9 \pm 55$ & $-15 \pm 9$ \\
    ~~~8594.48428 & $16620.0 \pm 59$ & $-21 \pm 9$ \\
    ~~~8595.51427 & $16643.6 \pm 78$ & $-4 \pm 11$ \\
    ~~~8554.64061 & $16607.3 \pm 63$ & $-26 \pm 10$ \\
    ~~~8555.64052 & $16619.1 \pm 56$ & $-13 \pm 9$ \\
    ~~~8574.59030 & $16639.2 \pm 82$ & $-25 \pm 12$ \\
    ~~~8556.66845 & $16606.4 \pm 67$ & $-8 \pm 10$ \\
    ~~~8559.66248 & $16587.7 \pm 64$ & $-3 \pm 10$ \\
    ~~~8572.59926 & $16608.6 \pm 64$ & $-12 \pm 10$ \\
    ~~~8570.59902 & $16604.7 \pm 64$ & $-15 \pm 10$ \\
    ~~~8566.59518 & $16607.8 \pm 55$ & $-23 \pm 9$ \\
    ~~~8569.61309 & $16595.1 \pm 56$ & $-27 \pm 9$ \\
    ~~~8568.58302 & $16603.2 \pm 56$ & $-25 \pm 9$ \\
    \midrule
    HARPS & & \\
    ~~~8578.56809 & $16640.7 \pm 3.0$ & $-23 \pm 4$ \\
    ~~~8834.81040 & $16655.1 \pm 2.0$ & $-18 \pm 2$ \\
    ~~~8832.79213 & $16632.6 \pm 2.3$ & $-17 \pm 3$ \\
    ~~~8823.83066 & $16646.5 \pm 2.0$ & $-18 \pm 2$ \\
    ~~~8806.82846 & $16664.0 \pm 2.0$ & $-13 \pm 2$ \\
    ~~~8802.78054 & $16647.0 \pm 2.0$ & $-16 \pm 2$ \\
    ~~~8840.72423 & $16643.0 \pm 2.0$ & $-14 \pm 2$ \\
    ~~~8579.56214 & $16641.8 \pm 2.0$ & $-21 \pm 2$ \\
    ~~~8580.55407 & $16641.4 \pm 2.0$ & $-18 \pm 2$ \\
    ~~~8836.78320 & $16661.0 \pm 2.0$ & $-14 \pm 2$ \\
    ~~~8581.53267 & $16637.4 \pm 2.0$ & $-17 \pm 2$ \\
    ~~~8582.58717 & $16647.7 \pm 2.0$ & $-13 \pm 2$ \\
    ~~~8583.55616 & $16656.9 \pm 2.0$ & $-14 \pm 2$ \\
    ~~~8576.59963 & $16651.3 \pm 3.4$ & $-13 \pm 4$ \\
    \midrule
    PFS & & \\
    ~~~8620.46359 & $-10.07 \pm 0.80$ & --- \\
    ~~~8766.87428 & $9.76 \pm 0.71$ & --- \\
    ~~~8761.88540 & $-10.21 \pm 0.78$ & --- \\
    ~~~8593.49024 & $8.57 \pm 1.07$ & --- \\
    ~~~8595.48683 & $15.47 \pm 1.18$ & --- \\
    ~~~8625.47898 & $15.25 \pm 1.24$ & --- \\
    ~~~8619.47176 & $-7.34 \pm 0.88$ & --- \\
    ~~~8623.47001 & $6.66 \pm 0.74$ & --- \\
    ~~~8591.54149 & $-6.09 \pm 0.99$ & --- \\
    ~~~8622.47493 & $-1.71 \pm 0.77$ & --- \\
    \bottomrule
    \end{tabular}
\end{table}

%RVs for target B
\begin{table}
    \centering
    \caption{Radial velocity data obtained for \plnameB .}
    \label{tbl:rvB}
    \begin{tabular}{lcc}
    \toprule
    \toprule
    BJD - 2450000 & Radial Velocity  (m/s) & Bisector (m/s) \\
    \midrule
    FEROS & & \\
    ~~~8543.64921 & $44723.7 \pm 8.7$ & $-4.0 \pm 12.0$ \\
    ~~~8546.65455 & $44692.3 \pm 10.7$ &   $-1.0 \pm 15.0$ \\
    ~~~8549.64390 & $44679.2 \pm 8.2$ &    $8.0 \pm 12.0$ \\
    ~~~8550.66827 & $44654.1 \pm 8.5$ &  $-12.0 \pm 12.0$ \\
    ~~~8551.59780 & $44691.6 \pm 9.5$ &  $-15.0 \pm 13.0$ \\
    ~~~8553.65940 & $44712.6 \pm 13.1$ &   $40.0 \pm 17.0$ \\
    ~~~8554.65929 & $44726.9 \pm 12.7$ &   $42.0 \pm 17.0$ \\
    ~~~8555.63065 & $44692.3 \pm 8.4$ &    $0.0 \pm 12.0$ \\
    ~~~8556.67850 & $44699.7 \pm 9.5$ &   $-7.0 \pm 13.0$ \\
    ~~~8558.55750 & $44685.1 \pm 10.1$ &   $33.0 \pm 14.0$ \\
    ~~~8560.66050 & $44639.7 \pm 8.8$ &  $-30.0 \pm 12.0$ \\
    ~~~8566.58573 & $44678.9 \pm 8.1$ &   $16.0 \pm 12.0$ \\
    ~~~8567.59465 & $44678.1 \pm 8.2$ &  $-32.0 \pm 12.0$ \\
    ~~~8571.64310 & $44723.8 \pm 10.7$ &   $15.0 \pm 15.0$ \\
    ~~~8572.60896 & $44653.8 \pm 10.6$ &    $7.0 \pm 15.0$ \\
    ~~~8574.58052 & $44665.1 \pm 12.1$ &   $23.0 \pm 16.0$ \\
    ~~~8593.48862 & $44730.5 \pm 11.3$ &  $-3.0 \pm 15.0$ \\
    ~~~8617.51392 & $44719.2 \pm 7.7$ &  $-20.0 \pm 11.0$ \\
    ~~~8619.48691 & $44681.9 \pm 9.0$ &    $0.0 \pm 13.0$ \\
    ~~~8623.46101 & $44665.8 \pm 16.2$ &    $7.0 \pm 20.0$ \\
    ~~~8624.47143 & $44700.0 \pm 12.8$ &   $33.0 \pm 17.0$ \\
    ~~~8592.58057 & $44689.6 \pm 18.3$ &  $-23.0 \pm 18.0$ \\
    ~~~8599.50975 & $44696.3 \pm 9.9$ &    $1.0 \pm 14.0$ \\
    ~~~8625.46358 & $44670.2 \pm 10.9$ &  $-31.0 \pm 15.0$ \\
    ~~~8626.47120 & $44696.6 \pm 10.2$ &  $-32.0 \pm 14.0$ \\
    ~~~8627.46390 & $44696.7 \pm 8.4$ &   $-6.0 \pm 12.0$ \\
    ~~~8637.49093 & $44685.7 \pm 7.4$ &    $4.0 \pm 11.0$ \\
    ~~~8638.46431 & $44718.8 \pm 9.4$ &  $-13.0 \pm 13.0$ \\
    ~~~8640.45795 & $44703.0 \pm 9.5$ &  $-13.0 \pm 13.0$ \\
    ~~~8641.47584 & $44700.3 \pm 13.1$ & $-128.0 \pm 17.0$ \\
    ~~~8642.47749 & $44737.6 \pm 11.4$ &  $-55.0 \pm 15.0$ \\
    ~~~8642.47684 & $44737.6 \pm 11.4$ &  $-55.0 \pm 15.0$ \\
    ~~~8643.46106 & $44699.1 \pm 9.1$ &    $9.0 \pm 13.0$ \\
    ~~~8645.47877 & $44772.1 \pm 12.0$ &    $7.0 \pm 16.0$ \\
    \midrule
    HARPS & & \\
    ~~~8836.75538 & $44730.1 \pm 3.9$ &   $-9.0 \pm 5.0$ \\
    ~~~8840.82178 & $44706.7 \pm 3.7$ &   $-1.0 \pm 5.0$ \\
    ~~~8851.72242 & $44726.2 \pm 3.7$ &  $-12.0 \pm 5.0$ \\
    ~~~8854.72962 & $44680.5 \pm 8.4$ &  $-35.0 \pm 11.0$ \\
    ~~~8876.71642 & $44732.5 \pm 4.8$ &   $-3.0 \pm 6.0$ \\
    ~~~8881.68161 & $44696.1 \pm 5.1$ &   $-5.0 \pm 7.0$ \\
    ~~~8885.61656 & $44725.6 \pm 7.2$ &    $1.0 \pm 9.0$ \\
    ~~~8890.66515 & $44715.4 \pm 4.8$ &  $-19.0 \pm 6.0$ \\
    ~~~8899.62785 & $44726.4 \pm 4.5$ &   $-6.0 \pm  6.0$ \\
    \midrule 
    CHIRON & & \\
    ~~~9181.80350 & $43294 \pm 24$ & --- \\
    ~~~9181.35804 & $43323 \pm 28$ & --- \\
    ~~~9184.77684 & $43328 \pm 35$ & --- \\
    ~~~9212.68165 & $43308 \pm 25$ & --- \\
    ~~~9213.69339 & $43256 \pm 39$ & --- \\
    ~~~9230.69126 & $43302 \pm 41$ & --- \\
    \bottomrule
    \end{tabular}
\end{table}

%--------------------------------------------------------------------

\begin{acknowledgements}
FR, RB, AJ and MH acknowledge support from project IC120009 ``Millennium Institute of Astrophysics (MAS)'' of the Millenium Science Initiative, Chilean Ministry of
Economy. RB acknowledges support from FONDECYT project 1241963. AJ acknowledges additional support from FONDECYT project 1171208. MTP acknowledges support by FONDECYT-ANID Post-doctoral fellowship no. 3210253. TT acknowledges support by the BNSF program "VIHREN-2021" project No. KP-06-DV/5.

The results reported herein benefited from collaborations and/or information exchange within NASA’s Nexus for Exoplanet System Science (NExSS) research coordination network sponsored by NASA’s Science Mission Directorate under Agreement No. 80NSSC21K0593 for the program ``Alien Earths".
% LCOGT
This work makes use of observations from the LCOGT network. Part of the LCOGT telescope time was granted by NOIRLab through the Mid-Scale Innovations Program (MSIP). MSIP is funded by NSF.
% TESS
KAC acknowledges support from the TESS mission via subaward s3449 from MIT.
%% KB
The postdoctoral fellowship of KB is funded by F.R.S.-FNRS grant T.0109.20 and by the Francqui Foundation.
% ExoFOP
This research has made use of the Exoplanet Follow-up Observation Program (ExoFOP; DOI: 10.26134/ExoFOP5) website, which is operated by the California Institute of Technology, under contract with the National Aeronautics and Space Administration under the Exoplanet Exploration Program.

This paper includes data collected with the TESS mission, obtained from the Mikulski Archive for Space Telescopes (MAST) data archive at the Space Telescope Science Institute (STScI). Funding for the TESS mission is provided by the NASA Explorer Program. STScI is operated by the Association of Universities for Research in Astronomy, Inc., under NASA contract NAS 5-26555.

We acknowledge the use of public TESS data from pipelines at the TESS Science Office and at the TESS Science Processing Operations Center.

Resources supporting this work were provided by the NASA High-End Computing (HEC) Program through the NASA Advanced Supercomputing (NAS) Division at Ames Research Center for the production of the SPOC data products.

Based on observations made with ESO Telescopes at the La Silla Paranal Observatory under program ID $0103.C-0442$ (PI RB).

This work makes use of observations from the Las Cumbres Observatory global telescope network.

\end{acknowledgements}

% WARNING
%-------------------------------------------------------------------
% Please note that we have included the references to the file aa.dem in
% order to compile it, but we ask you to:
%
% - use BibTeX with the regular commands:
%   \bibliographystyle{aa} % style aa.bst
%   \bibliography{Yourfile} % your references Yourfile.bib
%
% - join the .bib files when you upload your source files
%-------------------------------------------------------------------

\bibliographystyle{aa} % style aa.bst
\bibliography{tesscl,coauthors} % your references Yourfile.bib

\end{document}